\documentclass[a4paper,11pt]{article}
\usepackage{jheppub}
\pdfoutput=1

\usepackage{amsmath,amsfonts}
\usepackage{amssymb}
\usepackage{booktabs}
\usepackage{comment}
\usepackage{bbold}

\usepackage{float}
\usepackage{subfig}
\newcommand{\vol}{\operatorname{Vol}}
\newcommand{\prob}{\operatorname{prob}}

\graphicspath{ {figs/} }

\newcommand{\be}{\begin{equation}}
\newcommand{\ee}{\end{equation}}
\newcommand{\bea}{\begin{eqnarray}}
\newcommand{\eea}{\end{eqnarray}}

\newcommand{\nn}{\nonumber}

\newcommand{\one}{\mathbb{1}}

\renewcommand{\sl}{/\!\!\!}
\newcommand{\Dsl}{\,\sl\!D}

\newcommand{\mc}{\widetilde m}
\newcommand{\E}{\mathcal E}

\newcommand{\eps}{\epsilon}

\title{A Random Clockwork of Flavor}
\author[a]{Fernando Abreu de Souza}
\author[a]{and Gero von Gersdorff}
\affiliation[a]{Department of Physics, Pontif\'icia Universidade Cat\'olica do Rio de Janeiro, Rua Marqu\^es de S\~ao Vicente 225, Rio de Janeiro, Brazil}

\emailAdd{nando.abreu@poli.ufrj.br}
\emailAdd{gersdorff@puc-rio.br}

\abstract{We propose a simple clockwork model of flavor which successfully generates the Standard Model flavor hierarchies from random order-one couplings. With very few parameters we achieve distributions of models in excellent agreement with observation. 
We explain in some detail the interpretation of our mechanism as random localization of zero modes in theory space.
The scale of the vectorlike fermions is mostly constrained by lepton flavor violation with secondary constraints arising from rare meson decays.}

\begin{document}
\maketitle

\section{Introduction}

One of the striking features present in  the Standard Model (SM) of particle physics is the strongly hierarchical pattern of masses and mixings in the matter sector. 
On the one hand, different representations of the unbroken $SU(3)_c\times U(1)_{\rm EM}$ have noticeably different masses within a generation, in particular the masses of neutrinos are separated from those of the charged fermions by orders of magnitude. On the other hand, within each representation, there are strong inter-generational hierarchies, which are weakest for the neutrinos, and strongest for the up type quarks. A related hierarchy appears in the CKM matrix for the quarks, which parametrizes the mixing between the up and down quarks in the charged current interaction.

The Clockwork (CW) mechanism, originally formulated in \cite{Choi:2015fiu,Kaplan:2015fuy} to build consistent models of the Relaxion  \cite{Graham:2015cka}, has soon been realized to provide a general framework for constructing hierarchies in a natural way \cite{Giudice:2016yja}. 
In the flavor sector, it has been applied to explain the lightness of neutrino masses 
\cite{Ibarra:2017tju,Banerjee:2018grm,Hong:2019bki,Kitabayashi:2019qvi} as well as the  hierarchies in the charged fermion sector 
\cite{Burdman:2012sb,vonGersdorff:2017iym,Patel:2017pct,Alonso:2018bcg,Smolkovic:2019jow}.

In this paper, building on previous attempts \cite{vonGersdorff:2017iym}, we will provide an extremely simple model for the mass and CKM hierarchies which arise from  Lagrangian parameters of order one. Taking the latter randomly from  flat prior distributions, the resulting distributions for the eigenvalues and mixings depend  on five discrete parameters, the number of clockwork gears for each 
SM representation (two left handed doublets and three right handed singlets), as well as one continuous parameter of  $\mathcal O(10)$ whose  purpose is to suppress the bottom and tau masses with respect to the top mass. The  inter-generational hierarchies in contrast are  created from order-one random numbers. 
As we will see, the lightness of the neutrino masses cannot be convincingly explained in this simple model, so we will adopt the seesaw mechanism for this purpose, introducing one more non-stochastic parameter, the seesaw scale.

Our model can also be interpreted as spontaneously localizing SM fermion zero modes in the bulk of theory space, similar to Anderson-localization in condensed matter systems \cite{Craig:2017ppp}. We investigate in some detail the mechanism how this localization takes place.

The  scale that sets the masses of the CW gears is a priori a completely arbitrary parameter, which we will refer to as the CW scale.
We present an analysis of the main constraints on the CW scale from flavor changing neutral current (FCNC) observables, based on the dimension-six effective theory. Since there are only fermionic New Physics states, the sole tree-level FCNC operators are $\Delta F=1$ operators  
of the current-current type, $J^\mu_{\rm Higgs} J^\mu_{\rm fermion}$.
In particular, $\Delta F=2$ processes have to proceed via the exchange of a weak boson and hence are suppressed by the fourth power of the CW scale.
At the loop level, one can generate additional four-fermion and dipole type operators via loops of the Higgs and the new fermions.

This paper is organized as follows. In Sec.~\ref{sec:model} we introduce the model and explain its mechanism for the hierarchies. In Sec.~\ref{sec:CW} we explain the localization features of the zero mode in theory space and compare our random CW with the standard uniform model. In 
Sec~\ref{sec:fit} we present results of simulations and quantify the performance of the model for various choices of parameters.
Sec.~\ref{sec:dim6} and Sec.~\ref{sec:fcnc} are devoted to a discussion of the main FCNC effects in the effective field theory framework.
In Sec.~\ref{sec:conclusions} we present our conclusions. Two appendices deal with some technical details regarding the localization probabilities in theory space and the relation of the field basis used in the main text, and the mass eigenbasis.

\section{The model}
\label{sec:model}

\begin{figure}[htbp]
\begin{center}
\includegraphics[width=\linewidth]{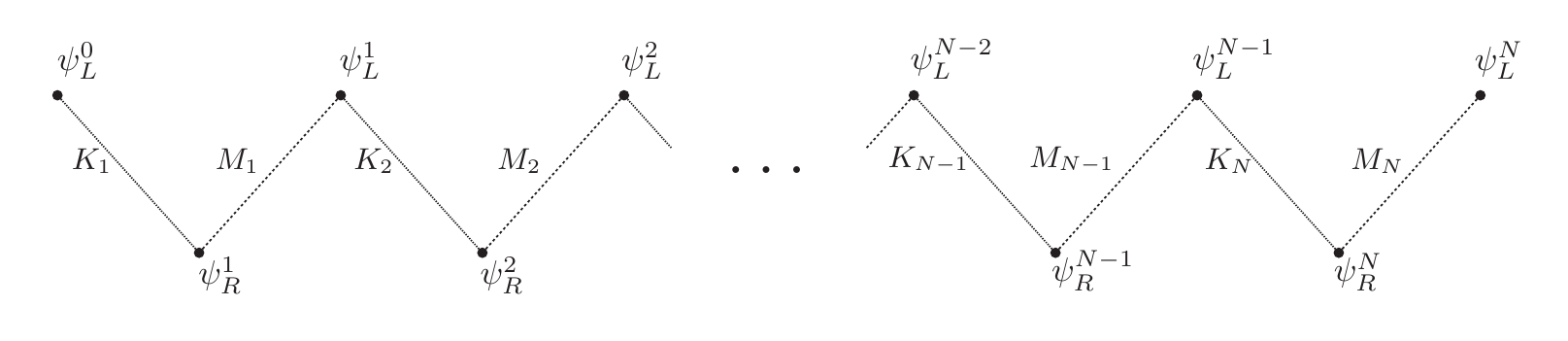}
\caption{The generic clockwork chain. There are three left-handed zero modes.}
\label{fig:chain}
\end{center}
\end{figure}

The model is defined by the following  Lagrangian
\be
\mathcal L=
\mathcal L_{q_L}+\mathcal L_{\ell_L}+\mathcal L_{u_R}+\mathcal L_{d_R}+\mathcal L_{e_R}
-( \bar q^0_LH C_d d^0_R+  \bar q^0_{L}\tilde H C_u u^0_{R}+ \bar \ell_L^0 H C_e e^0_R +h.c.)\,,
\label{eq:model}
\ee
where the bilinear part of the Lagrangian is given in terms of 
\be
\mathcal L_{\psi_L}=\sum_{i=0}^{N_\psi}i\bar \psi_L^{i}\,\sl{\! D}\psi^{i}_L+
\sum_{i=1}^{N_\psi}i \bar \psi_R^i\,\sl{\! D}\, \psi_R^i-\sum_{i=1}^{N_\psi}\bigl[\bar \psi^i_R M^{\psi}_{i} \psi_L^{i}-\bar \psi^i_R K^{\psi}_{i} \psi_L^{i-1}+h.c.\bigr]\,,
\label{eq:modelbis}
\ee
and the same for $\mathcal L_{\psi_R}$ with $L\to R$ everywhere.
All fermions carry implicit generation indices. A cartoon of one of the five sectors is shown in Fig.~\ref{fig:chain}. We will refer to the points in theory space labeled by $i$ as sites, and additionally we will refer to the first and last sites as the "boundaries" of theory space.
The matrices $M^\psi_i$ and $K^\psi_i$ are complex $3\times 3$ matrices with dimension of mass.
The last term in Eq.~(\ref{eq:model}) contains the coupling of the CW gears to the SM Higgs, where the "proto Yukawa couplings" $C_{d,u,e}$ are  complex, dimensionless matrices.
Notice that the $q$ and $\ell$ fields have one more left handed than right handed field per generation, while for the $u$, $d$ and $e$ fields it is the other way around. Therefore, we are guaranteed to have three left handed (right handed) zero modes for $q$ and $\ell$ ($u$, $d$ and $e$), which are identified with the corresponding SM fields. These zero modes will be linear combinations of all CW gears. However, as we will see, it is very convenient to integrate out the CW gears without going to the mass eigenbasis and use the $\psi^0$, i.e., the boundary fields,  as "interpolating fields" for the zero modes.~\footnote{This is similar to holography in extra dimensions, where one works with the brane value of bulk fields as opposed to the true zero mode}

Let us briefly discuss the symmetries of this Lagrangian. Ignoring the coupling to the Higgs, the sectors decouple. Then, say, in the $q$ sector, one has an $U(3)_R^{N_q}\times U(3)_L^{N_q+1}$ under which the $M^q_{i}$ and $K^{q}_i$ transform as $({\bf  3}_{R,i},{\bf \bar 3}_{L,i})$ and $({\bf  3}_{R,i},{\bf \bar 3}_{L,i-1})$ spurions respectively.
Similar symmetries appear for the $d$, $u$, $e$, and $\ell$ sectors. In order to guarantee that only nearest neighbours couple, it is sufficient to invoke an Abelian subgroup of this big symmetry.
Notice however that there is no apparent symmetry that would enforce the universality (site-independence) of the matrices $K^i$ and $M^i$. \footnote{The only obvious way would be a "translational invariance" which is however necessarily broken by the boundaries.} 
There is one more symmetry that is worth mentioning. It is the discrete transformation
\bea
q_L^k\to q_L^{N_\psi-k}\,,\qquad q_R^k\to q_R^{N_q+1-k}\,,\qquad M^q_{k} \to K^q_{N_u+1-k}\,,\qquad K^q_{k} \to M^q_{N_u+1-k}\,,
\eea
which corresponds to a reflection of the diagram in Fig.~\ref{fig:chain} around its center. Below we will make use of the shorthands
\be
Q^q_k\equiv (M^q_k)^{-1}K^q_k\,,
\label{eq:qdef}
\ee
(and similarly for the other fields $\psi=u,d,e,\ell$)
which transform as
\be
Q^q_{k}\to (Q^q_{N_q+1-k})^{-1}\,.
\ee

Solving the equations of motion for the vectorlike clockwork gears leads to the following  Lagrangians for the interpolating fields $\psi^0$:
\be
\mathcal L_{\psi_L}= i\bar\psi^0_L\, \Pi_{\psi^0}\Dsl\psi_L^0\,,
\ee
where the form factors $\Pi_{\psi}(p^2)$ are determined from the following recursion relation
\be
\Pi_{\psi^{\ell-1}}=1+(Q^\psi_{\ell})^\dagger \frac{1}{\Pi_{\psi^{\ell}}^{-1}+|M^\psi_\ell|^{-2} \Dsl^2}Q^\psi_\ell\,,\qquad \Pi_{\psi^N}=1\,,
\ee
with $|M|^2\equiv M^\dagger M$ and $Q\equiv M^{-1}K$.
We stress that no approximations have been made (except for the fact that we work at tree level for now) and hence the Lagrangians are exact even though they  contain an infinite number of derivatives. 
For momenta small compared to the lightest gear, we expand this to quadratic order in  derivatives 
\be
\Pi_{\psi^0}\approx Z_{\psi^0}- A_{\psi^0} \Dsl^2\,.
\label{eq:ffexp}
\ee
As we will see, the $Z_{\psi^0}$ are hierarchical matrices with eigenvalues $>1$ which will lead to hierarchical Yukawa couplings after canonical normalization. The matrices $A$ on the other hand will describe the flavour violating dimension-six operators, and are evaluated in Sec.~\ref{sec:ops}. One easily obtains the recursion relations
\be
Z_{\psi^{\ell-1}}=1+(Q^\psi_\ell)^\dagger Z_{\psi^\ell}Q^\psi_\ell\,, \qquad Z_{\psi^{N}}=1\,,
\label{eq:recZ}
\ee
which we can solve explicitly as
\be
Z_{\psi^{\ell-1}}=1+ (Q^\psi_\ell)^\dagger Q_\ell^\psi+ (Q^\psi_{\ell+1}Q^\psi_\ell)^\dagger Q_{\ell+1}^\psi Q_\ell^\psi+\dots+(Q^\psi_N \cdots Q^\psi_\ell)^\dagger (Q^\psi_N \cdots Q^\psi_\ell)\,.
\label{eq:Z}
\ee
As shown below, even for order-one random matrices $K$ and $M$, 
 the matrix $Z$ is very hierarchical, with eigenvalues ranging from $z\gtrsim 1$ to $z\gg 1$. 

After canonical normalization
\be
\psi=(Z_{\psi^0})^{\frac{1}{2}}\psi^0\,,
\ee
we obtain the physical Yukawa couplings
\be
Y_u=\E_{q} C_u \E_u\,,\qquad Y_d=\E_q C_d \E_d \,,\qquad Y_e=\E_\ell C_e \E_e\,,
\label{eq:Yuk}
\ee
where 
\be
\E_\psi\equiv (Z_{\psi^0})^{-\frac{1}{2}}\,.
\label{eq:Eps}
\ee

The model we are considering is essentially the one studied previously in Ref.\cite{vonGersdorff:2017iym}. Compared to Ref.\cite{vonGersdorff:2017iym}, we are considering 
only one chirality when coupling the neighbouring sites. 
In Ref.\cite{Patel:2017pct} a related model was proposed in which uniform  (site independent) Hermitian random matrices $Q_i=Q$   
were assumed.
In the model of Ref.~\cite{Alonso:2018bcg}, clockwork chains of different lengths within each fermion species were considered, with site-independent couplings.
Recently, the authors of Ref.~\cite{Smolkovic:2019jow} considered classes of models which in some cases are very similar to our model, in particular, the couplings between the clockwork fields were taken site-dependent and random. 
As pointed out, the assumption of uniformity (that is, site-independent couplings) is not easy to justify by four-dimensional symmetry principles. However, it is also  unnecessary, as  a successful model of flavor arises just from our Lagrangian, without any further assumptions.

In the rest of this section we will review how our mechanism generates hierarchical Yukawa couplings and CKM angles \cite{vonGersdorff:2017iym}. 
In the following we will call a complex matrix $B$ hierarchical if its singular values (the square roots  of the eigenvalues of $BB^\dagger$) are hierarchical, $b_1\ll b_2\ll b_3$. 
We will work under the assumption that the matrices $K^\psi_i$ and $M_i^\psi$ have  order-one complex entries in units of some "clockwork scale"  that we leave unspecified for now.  Notice that the matrix $Q_i^\psi=(M_i^\psi)^{-1}K_i^\psi$ and hence the Yukawa couplings are independent of this scale, which will thus show up only in higher dimensional operators.
 We will see that for any given order-one random matrices $K$ and $M$  it is very likely that the matrix $Z$ is very hierarchical with eigenvalues ranging from unity to very large values, and  this hierarchy is inherited by the Yukawa couplings in Eq.~(\ref{eq:Yuk}) after canonical normalization.

The mechanism is based on the following two observations \cite{vonGersdorff:2017iym}
\begin{enumerate}
\item
{\it Hierarchy from products}.
The product of randomly chosen order-one  matrices  quickly becomes  hierarchical with increasing number of factors.
\item
{\it Common factor alignment}.
For $B$ a hierarchical complex matrix, and $C$ and $D$ randomly chosen complex matrices,  $BC$ and $BD$ are likely to be left-aligned, i.e.~the eigenvectors of the Hermitian matrices $BC(BC)^\dagger $ and $BD(BD)^\dagger$ are approximately aligned.
\end{enumerate}

\begin{figure}[!htb]%
    \centering
    \subfloat[Basic distribution]{{\includegraphics[width=7cm]{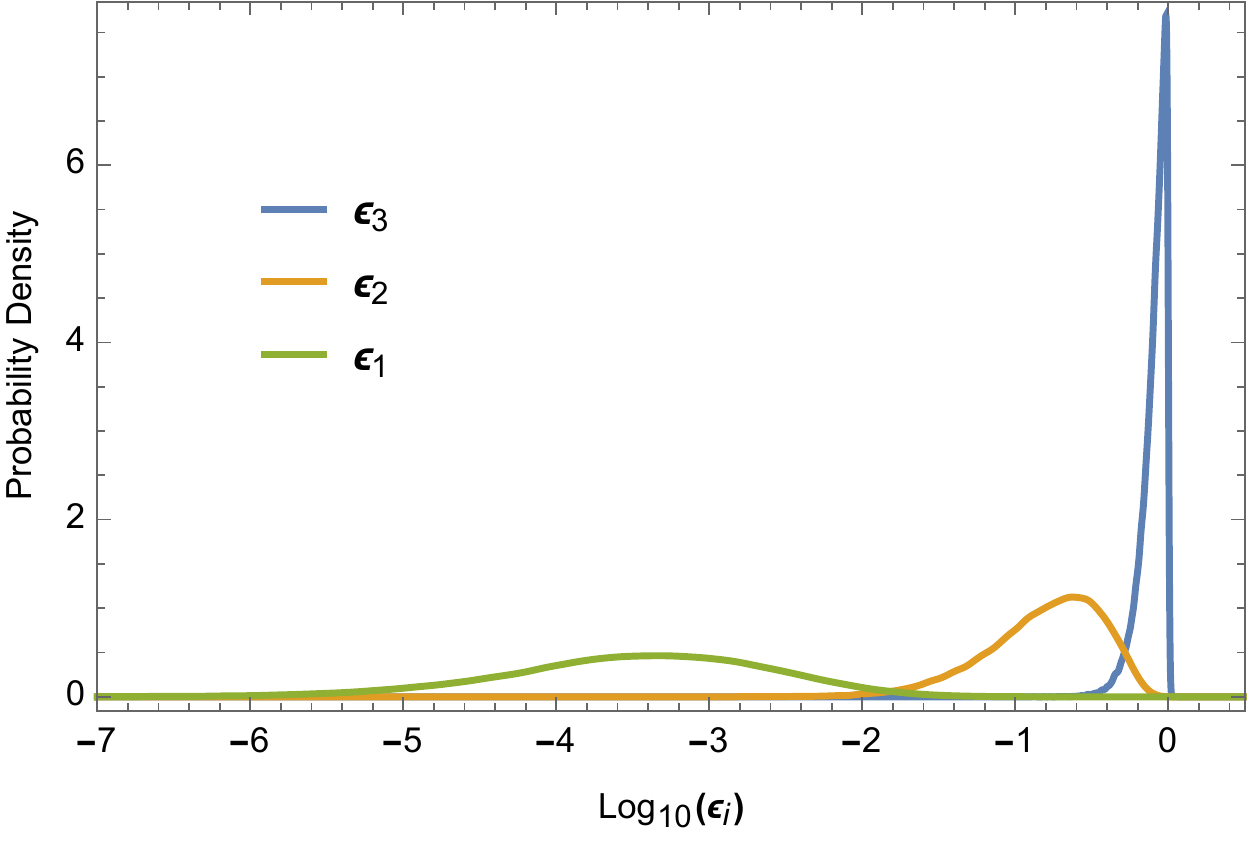}}\label{fig:epsdist}}%
    \quad
    \subfloat[htc][Modified distribution]{{\includegraphics[width=7cm]{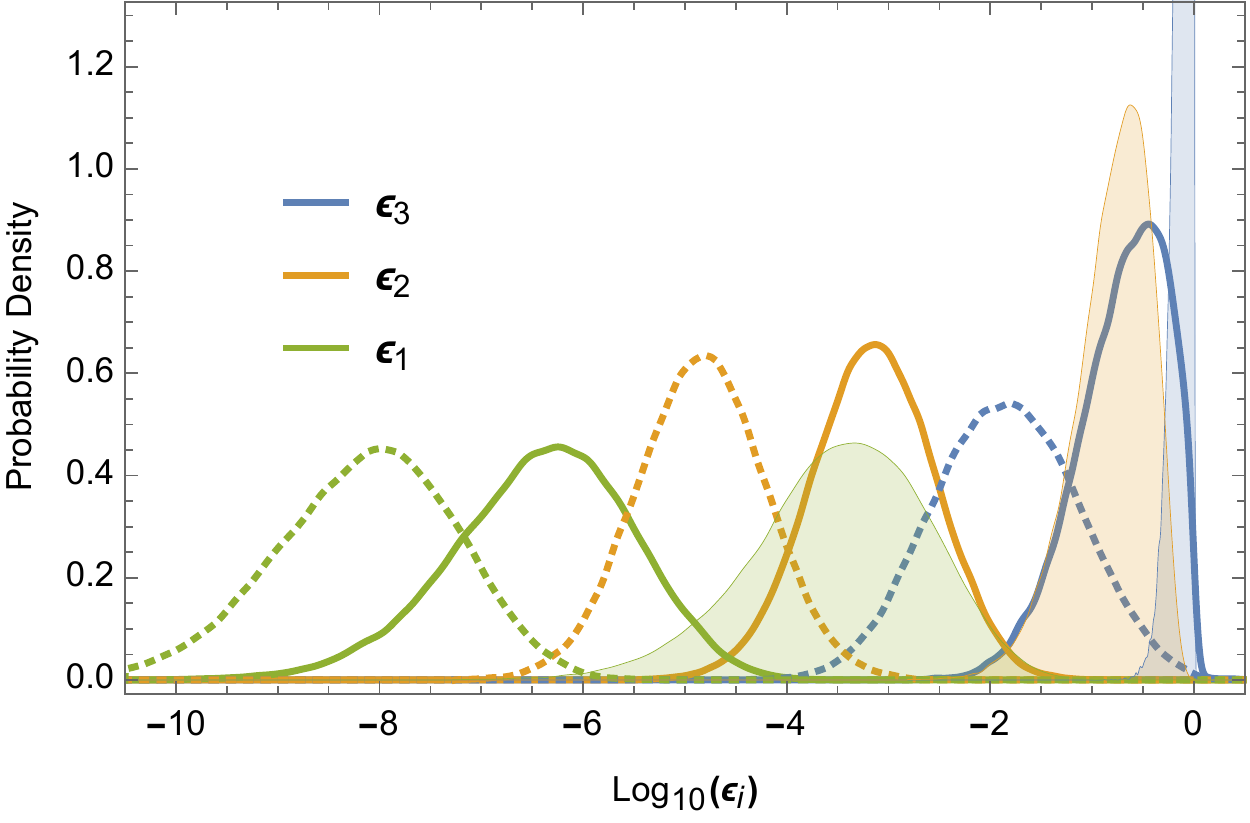}}\label{fig:epsdistb}}%
    \caption{(a): Distribution of eigenvalues of the matrix $\E$ with $N=10$ clockwork gears. (b): Modified distribution with $\xi=2$ (solid) and $\xi=3$ (dashed), see text for details. For comparison we show again the unmodified distribution (shaded).}
\end{figure}

The first observation guarantees that the $Z$ matrices  (and hence the $\E$ matrices) are very hierarchical. Specifically, the second term in the expression for $Z_{\psi^{\ell-1}}$, Eq.~(\ref{eq:recZ}) is more hierarchical than $Z_{\psi^\ell}$ because of the additional factors of $Q^\psi_\ell$. Adding the identity to it simply raises each of the three eigenvalues of the second term by one.\footnote{Notice that this guarantees also that all eigenvalues are larger than one.} This slightly mitigates the hierarchy, in particular it "resets" possible eigenvalues $\ll 1$ to one, while it essentially does not affect very large eigenvalues. 
Typical distributions for the eigenvalues of $\E$ are shown in Fig.~\ref{fig:epsdist}, where we use uniform priors with $\left|\operatorname{Re}(M_i)_{kl}\right|<1$, $\left|\operatorname{Im}(M_i)_{kl}\right|<1$ and analogously for $K_i$. These distributions depend on only one discrete free parameter, the number of clockwork gears $N$. A characteristic feature is that the largest eigenvalue is very sharply peaked just below $\eps=1$, while the distributions for the smaller eigenvalues become more and more spread out.

The second observation above guarantees that the CKM matrix is near-diagonal. As is evident from Eq.~(\ref{eq:Yuk}), the Yukawa couplings for up and down sector have the common hierarchical  factor $B=\E_{q}$, which guarantees the alignment of the left handed rotations $V_u$ and $V_d$ that appear in the singular value decomposition for the quark Yukawa couplings. Notice that the rotation matrices themselves are not necessarily  close to the identity, only the combination $V_{\rm CKM}=(V_u)^\dagger V_d$.

A more explicit way of seeing these features is to go to the basis in which the matrices $\E_\psi$ are diagonal. This is achieved by order-one, gauge-invariant, unitary rotations, such that up to order one-numbers, the Yukawa couplings become $(Y_u)_{ij}\sim \eps_{q^i}\eps_{u^j}$ etc. This in turn leads to eigenvalues and mixing matrices 
\be
y_{u^i}\sim  \eps_{q^i}\eps_{u^i}\,,\qquad y_{d^i}\sim  \eps_{q^i}\eps_{d^i}\,,\qquad (V_{\rm CKM})^{ij}\sim \eps_{q^i}/\eps_{q^j}\,,
\label{eq:approxevmix}
\ee
with $i<j$ in the last relation. 
A rigorous and systematic treatment underlying the  behaviour shown in Eq.~(\ref{eq:approxevmix}) has recently been given in \cite{vonGersdorff:2019gle}.

So far, the distributions for the Yukawa couplings depend solely on five discrete parameters $N_\psi$. As it turns out, these distributions are too rigid for the following reason. As can be seen for  from Fig.~\ref{fig:epsdist}, the largest eigenvalue of the matrix $\E_\psi$ is always of order one. This introduces a problem for the down quark and charged lepton sectors, which require third generation Yukawa couplings of the order of $10^{-2}$. There are two simple solutions to this problem. The first one is to include an overall suppression factor in front of the proto-Yukawa couplings for the down and lepton sectors, which could for instance be due to a large $\tan\beta$ in a  supersymmetric or a two Higgs doublet (2HDM) version of our model.
The second possibility is to suppose that the $K^\psi$ matrices  are generated at a slightly different scale than the $M^\psi$ matrices. Denoting the ratio of such scales by $\xi_\psi$, one can conveniently incorporate this effect by writing the corresponding $Q$ matrices as
\be
Q^\psi=\xi_\psi (\hat M^\psi)^{-1}\hat K^\psi\,,
\label{eq:xi}
\ee
where the hatted quantities are dimensionless complex matrices taken to be of order one.
For $\xi\gg 1$, one can approximate
\be
(\E_{\psi})^2\approx (Q^\psi_N \cdots Q^\psi_\ell)^{-1} (Q^\psi_N \cdots Q^\psi_\ell)^{\dagger-1}\,.
\ee
The eigenvalues of this matrix are expected to shift to smaller values. This can be seen for instance for the cases $\xi_\psi=2,3$ explicitly shown in Fig.~\ref{fig:epsdistb}. It can also be observed that this increases the hierarchy middle and largest eigenvalues, while the hierarchy between smallest and middle eigenvalue is unaffected.\footnote{Also notice that in this limit one could work with simpler non-Hermitian matrices $\E_\psi=(Q^\psi_N \cdots Q^\psi_\ell)^{-1}$ when normalizing canonically. The two choices differ by a unitary matrix.
Whatever the choice, the structure of the $\E_\psi$ matrices is now a simple product, showing once more that our mechanism is a consequence of the "hierarchy from products" property described above.}
For the opposite case, $\xi_\psi\ll1$, all three eigenvalues of $\E_\psi$ will be very close to one, and the hierarchies disappear altogether. 

Finally, we must accommodate neutrino masses. Within our paradigm it is difficult to use the clockwork mechanism to explain the smallness of neutrino masses (for instance by introducing $N_\nu$ sterile clockwork gears with $\xi_\nu>1$), as a large number of gears will necessarily mean a large hierarchy between the neutrinos themselves, contrary to observation. We will thus simply implement the see-saw mechanism, or, equivalently write the low-energy Weinberg operator
\be
\mathcal L_W=-\frac{1}{2}\bar \ell_L^0\tilde H \, (C_\nu M_{\nu_R}^{-1} C_\nu^T)     \tilde H^T\ell_L^{0c}+{\rm h.c.}
\ee
We will separate the right handed mass matrix into a scale $m_{\nu_R}$ and a dimensionless symmetric complex matrix $\hat M_{\nu_R}$ that we will take as random order-one numbers
\be M_{\nu_R}=m_{\nu_R}\hat M_{\nu_R}.\ee
The seesaw scale $m_{\nu_R}$ will be another free (non-stochastic) parameter of the theory.

\section{Clockwork Lagrangians: uniform versus random}
\label{sec:CW}

Up to now we have not really made any connection to the usual CW paradigm of "coupling suppression from zero mode localization". Moreover,  our CW Lagrangian 
does not even seem to contain any small order parameter $q$ controlling these suppressions. In this section we will clarify these issues and also comment on an  interesting observation made recently regarding a link to Anderson localization in condensed matter systems \cite{Craig:2017ppp}. This section is not essential for the understanding of the rest of the paper, Sec.~\ref{sec:fit}, \ref{sec:dim6} and \ref{sec:fcnc}.
 
We start out by considering a single generation of fermions, in preparation for the realistic three-generation case to be presented after.
The Lagrangian reads:
\be
\mathcal L=
\sum_{i=0}^{N}i\bar \psi_L^{i}\,\sl{\! D}\psi^{i}_L+
\sum_{i=1}^{N}i \bar \psi_R^i\,\sl{\! D}\, \psi_R^i-\sum_{i=1}^{N}\bigl[\bar \psi^i_R m_i \psi_L^{i}-\bar \psi^i_R k_i \psi_L^{i-1}+h.c.\bigr]
-( \bar{\mathcal  O} \psi_L^0+h.c.)\,,
\label{eq:cw1}
\ee
where we have written a fermionic source $\mathcal O$ to keep track of couplings to the other sectors of the theory.
Reading off the mass matrix for the fermions gives 
\be
\mathcal M_{ij}=m_i\delta_{i,j}-k_i\delta_{i,j+1}\,,\qquad 1\leq i\leq  N\,,\quad  0\leq j\leq  N\,,
\ee
which has a left-handed zero mode  $\psi_L\equiv f^*_i\psi_L^i$ with  
\be
f_i\equiv \eps\, q_1q_2\cdots q_i\,,
\qquad
 q_j\equiv \frac{k_j}{m_j}\,,
 \label{eq:wf0}
\ee
and
\be
\eps\equiv (1+|q_1|^2+|q_1q_2|^2+\dots+|q_1q_2\cdots q_n|^2)^{-\frac{1}{2}}\,.
\label{eq:eps}
\ee 
The eigenvector $f_i$ will be called the zero mode wave function in what follows. Note that the zero mode wave function at site zero is simply $f_0=\eps$. 

In the standard CW scenario one  choses the $q_i$ uniformly, $q_i=q$. We will refer to this scenario as the "ordered" or "uniform"  CW model.
On the other hand, taking the $q_i$ different (and random) at each point, will result in  the "disordered" or "random" CW model.
In the standard uniform case, the wave function 
\be
f_i=\eps (q)^i\,,
\ee
is monotonically decreasing (for $|q|< 1$) or monotonically increasing (for $|q|>1$). In the former case, one has $\eps\approx1$ and the coupling of the zero mode to the operator $\mathcal O$ is unsuppressed. In the second case, one has $\eps\ll1$ and the coupling of the zero mode is suppressed by $\eps$. 
Hence, unless $q=1$, the zero-mode wave function is strongly localized at either boundary of the lattice. 
Moreover, the wave functions for the gears can be computed analytically and are essentially delocalized over the lattice \cite{Alonso:2018bcg}. 

On the other hand, for the random CW model the zero mode can peak at any site. However, as we will show, the localization is still rather narrow around that site, such that the wave function $f_0$ is typically still suppressed.
In App.~\ref{sec:anderson} it is shown that for randomly chosen parameters $m_i$ and $k_i$, the probability $p_{N,j}$ for the maximum of the zero mode wave function to occur at site $j$ is given by 
\be
p_{N,j}=p_{j,0}p_{N-j,0}\,,\qquad p_{j,0}=\frac{(2j-1)!!}{(2j)!!}\,.
\label{eq:Pmain}
\ee
To derive this result it is assumed that the functional form of the prior distributions for $k_i$ and $m_i$ are the same and site-independent. Except for this assumption, the localization probabilities in Eq.~(\ref{eq:Pmain}) are independent of the chosen prior. \footnote{
By introducing an asymmetry into the distributions  for the $k_i$ and $m_i$ parameters (that is, by making the $m_i$ systematically larger or smaller than the $k_i$) one generates a bias in the localization probabilities. This is equivalent to introducing the parameter $\xi$ in Eq.~(\ref{eq:xi}). Then for $\xi>1$ the wave function is expected to be more likely to peak at larger values of $j$, while for $\xi>1$ they the peaks are pushed towards smaller values of $j$.}

\begin{figure}
\centering
\includegraphics[width=8cm]{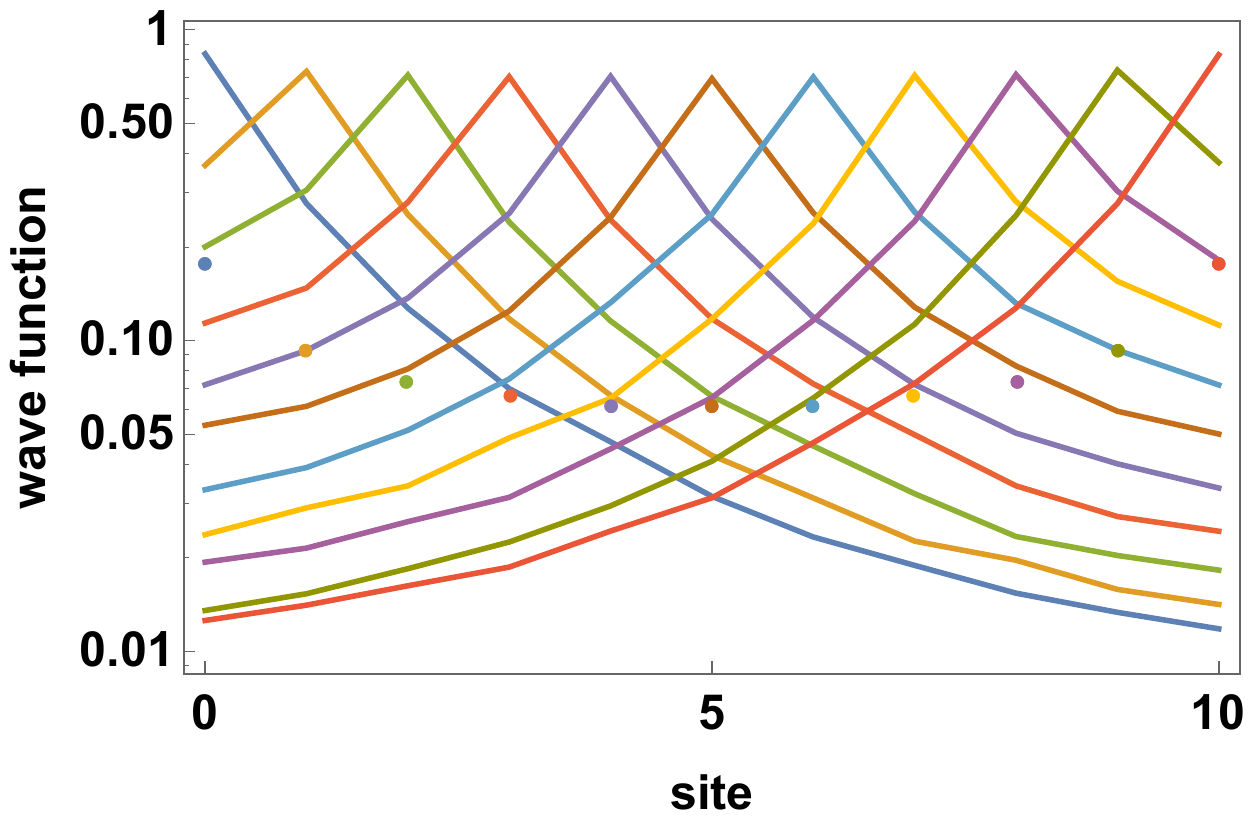}
\caption{The averaged zero mode wave functions for a random CW with $N=10$. We separately average over the wave functions that peak at a given site. The dots mark the probabilities for the peaks to occur at that site, given in Eq.~(\ref{eq:Pmain})}
\label{fig:wfave}
\end{figure}

The next step is to see how localized the wave function becomes, given that it is peaked at site $j$. 
This depends on the prior.
For identical  priors for $m$ and $k$,
it is intuitively clear (and easily shown), that the distribution for $x=\log |q|^2$ is then symmetric under $x\to-x$.
For instance, for uniform priors it is simply given by $p(x)=\frac{1}{2}e^{-|x|}$.
We show in Fig.~\ref{fig:wfave}  the averaged wave functions for the case of $N=10$.\footnote{Since the distributions for the wave functions are rather asymmetric, we chose to represent the typical wave function by its median instead of its mean.}
The values of $\eps$ correspond to the left end points of the curves. 
Naively one might have expected that the wave function Eq.~(\ref{eq:wf0}) with randomly chosen coefficients $q_i$ is uniformly spread over the lattice. Instead, one finds that they are rather strongly localized around their maxima.

A semi-quantitative understanding of the localization mechanism can be obtained as follows. 
Let us consider the cases with maximum at fixed site $j$ (with probabilities given above).
We then consider the (conditional) probability that the wave function decreases when we move away from the maximum, say from site
 $j+k-1$ to site $j+k$ with $k=1\dots N-j$.
\be
\prob\bigl(|f_{j+k}|< |f_{j+k-1}|\,\bigm |f_{\rm max}=|f_j|\bigr)\equiv p_-(k)\,.
\ee
Clearly $p_-(1)=1$ (otherwise $|f_j|$ is not maximal). In App.~\ref{sec:anderson} it is shown that $p_-(k)$ is monotonically decreasing with $k$, but it always stays above $\frac{1}{2}$, with the universal (prior-independent) limiting value
\be
p_-(N-j)=\frac{(N-j)}{2(N-j)-1}>\frac{1}{2}.
\ee
Hence there is a strong bias for the wave function to decrease to the right of the peak, which weakens but never goes away once we move to larger $k$. 
A completely analogous argument applies to the left of the peak. This fact explains the shapes in Fig.~\ref{fig:wfave}.
In summary, the probability for the wave function to rapidly decay  away from its maximum is large.

A further comment concerns the
massive modes or "clockwork gears". In contrast to the uniform case they are also localized very sharply at individual sites. It is thus typically the case that only very few heavy modes couple to the source $\mathcal O$.

In Ref.~\cite{Craig:2017ppp} a similar class of models was studied, and it was pointed out that the localization  features observed here are closely related to an effect that is known as Anderson localization in condensed matter systems~\cite{Anderson:1958vr}. To see this parallel here, consider the Hermitian matrix\footnote{Similar considerations apply to $H_R=\mathcal M\mathcal M^\dagger$.} $H_L=\mathcal M^\dagger\mathcal M$ 
\be
H_L=E_i\delta_{ij}+V^*_j\delta_{i+1,j}+V_i\delta_{i,j+1}\qquad (i,j=0\dots N)\,,
\ee
with
$E_i\equiv |k_{i+1}|^2+|m_i|^2$ and $V_i\equiv -m^*_i k_i$.
 This matrix takes the form of a Hamiltonian over some  discrete (one-dimensional) lattice with nearest neighbour interactions, known as the  Anderson tight binding model. 
 This model is known to localize its energy eigenstates similarly to our model, implying that  any localized wave functions only have overlap with very few energy eigenstates and hence do not diffuse over time \cite{Anderson:1958vr}.
The original models only considered the $E_i$ to be stochastic and the interactions $V_i$ to be fixed (and uniform), but subsequent studies have shown that localization also occurs for off-diagonal disorder \cite{PhysRevB.24.5698}.

\begin{figure}[htbp]
\begin{center}
\includegraphics[width=8cm]{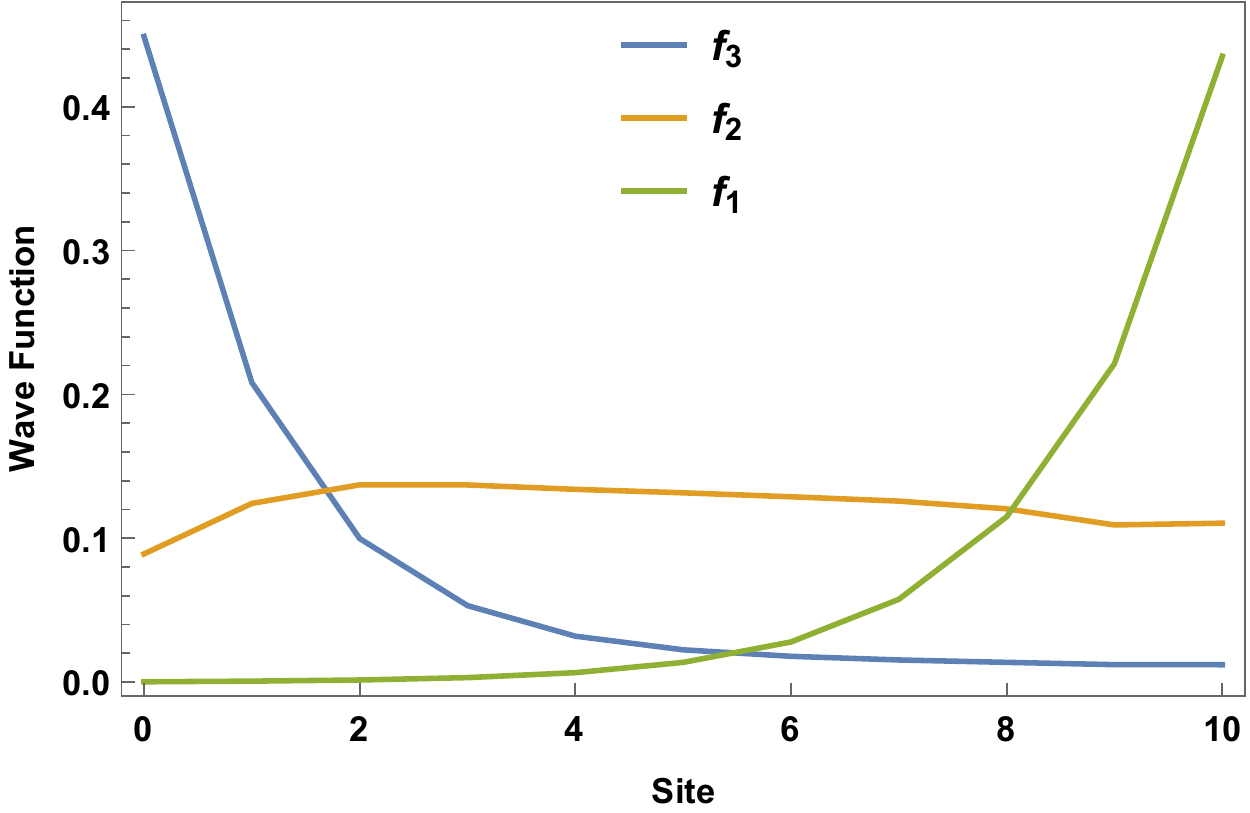}
\caption{Average localization of the three zero mode wave functions in theory space. We display the medians of the wave functions $f_i$ defined to be the three columns of Eq.~(\ref{eq:wfs}). }
\label{fig:loc3g}
\end{center}
\end{figure}

Let us now move to the case of three generations. There are now three zero modes and their orthonormal wave functions read
\be
F'=\begin{pmatrix}
1\\Q_1\\Q_2 Q_1\\\vdots\\
Q_N\cdots Q_1
\end{pmatrix}\E\,.
\ee
$F'$ is a $3(N+1)\times 3$ matrix whose three columns are the three zero mode wave functions and $\E$ is the same matrix encountered in Eq.~(\ref{eq:Eps}). Notice that each site has three fields living on it, and accordingly each wave function has three components per site.
Diagonalizing  $\E =V\E^{\rm diag}V^\dagger$, we find
\be
F'=\begin{pmatrix}
1\\Q_1\\Q_2 Q_1\\\vdots\\
Q_N\cdots Q_1
\end{pmatrix}V\E^{\rm diag}V^\dagger\,,
\label{eq:wfs}
\ee
with $V$ unitary. One can absorb the factor of $V^\dagger$ on the right by performing a further rotation of the zero mode fields with $V$.
In this basis the zero mode wave functions read $F=F'V$ and 
  their site-zero components  are given by the eigenvalues $\eps_i$ of $\E$.
  As in the case of one generation, the $\eps_i$ parametrize the mixing of the zeros modes with the boundary fields.
It is then tempting to speculate that the smaller the value of $\eps_i$ the further  the $i^{\rm th}$ zero mode wave function is localized away from site zero.
This can be easily verified in a numerical simulation. In Fig.~\ref{fig:loc3g} we show the average wave function of each of the three zero modes for a clockwork chain with $N=10$. 
Notice that the  three left endpoints of the curves correspond precisely to the medians of the distributions for the $\eps_i$ shown in Fig.~\ref{fig:epsdist}.\footnote{For $i=1$ ($i=3$) we find that 85\%  (90\%) of the wave functions peak at site $10$ (site 0). For the middle eigenvalue, the probability to peak at a given site is roughly site-independent, and we get a similar behaviour as in the one-generation case. This behaviour is not visible in Fig~\ref{fig:loc3g}, which only shows the average wave function independent of the peak location, in contrast to Fig.~\ref{fig:wfave}. } The correlation between localization and $\epsilon$ suppression is clearly visible.

In summary, in this section we have shown that our model can be interpreted as creating a spontaneous separation of the three zero modes of each fermion species in theory space. One zero mode is likely to live near site zero (where the Higgs is located), one is likely to live near the  opposite boundary, and yet another one is likely to live in the bulk. These modes are then identified, respectively, as the third, first and second generation of each species.

\section{Simulation}

\label{sec:fit}
\subsection{Quarks}

In order to assess how well the proposed model can naturally reproduce the observed strong hierarchical patterns in the fermion sector, one can perform 
 simulations to obtain the distributions of Yukawa eigenvalues, as given by Eq. (\ref{eq:Yuk}), and the related mixing angles. To this end, the  "proto-Yukawas" $C_{u,d}$, as well as the set of dimensionless matrices $\hat M^{\psi}_k$ and $\hat K^{\psi}_k$, were taken as random order one, $3 \times 3$ complex matrices.~\footnote{For simplicity we chose uniform priors $p(z)=\frac{1}{4}$ in the square $|\!\operatorname{Re}(z)|<1$, $|\!\operatorname{Im}(z)|<1$.} Besides the discrete parameters $N_{q}$, $N_{u}$ and $N_{d}$, there are three continuous parameters, $\xi_{q}$, $\xi_{u}$ and $\xi_{d}$  that we introduced in Eq.~(\ref{eq:xi}). 
Our task is now to optimize these six parameters and achieve the most favorable distribution for the observed quark masses and mixings.

In order to measure how well the distribution performs we proceed as follows. First, from the data of the simulation for a given choice of parameters, we compute the mean values $\mu_i$ as well as the variances $\sigma_i$ and the correlation matrix $C_{ij}$  of the eigenvalues and CKM angles. This defines a Gaussian approximation for the theoretical distribution of models. More precisely, we opt to work with the logarithms of the observables, 
\be
\{x_i\}=\log\{\theta_{12},\theta_{23},\theta_{13},J,y_u,y_c,y_t, y_{d},y_s,y_b\},\qquad 1\leq i\leq  10\,,
\ee
as the resulting distributions are much more symmetric than the variables themselves, and hence the Gaussian approximation  in terms of the first two moments is better than in the linear case.
We then define a $\chi^2$ function 
\be
\chi^2(x_i)=\sum_{i,j} \frac{x_i-\mu_i}{\sigma_i}(C^{-1})_{ij}\frac{x_j-\mu_j}{\sigma_j}\,,
\ee
 evaluate it at the experimental central values $\chi^2(x_i^{\rm exp})$, and optimize the parameters to obtain a minimal $\chi^2$.
For convenience, we summarize in Tab.~\ref{tab:exp} the experimental data.

Notice that this procedure is in a way the exact opposite to the standard fit of a theoretical model to experimental data. There, the parameters of the model are adjusted in order to minimize $\chi_{\rm exp}^2(x^{\rm th}_i)$. Here, the parameters  of the theoretical distributions are adjusted in order to minimize 
$\chi^2_{\rm th}(x_i^{\rm exp})$.

The distributions for the CKM angles to a good approximation only depend on $N_q$ (and weakly on $\xi_q$), while the up and down type mass eigenvalues only depend on the subsets $\{N_q, \xi_q, N_u,\xi_u\}$ and $\{N_q, \xi_q, N_d,\xi_d\}$ respectively.

There are a variety of effects that determine the behaviour of the $\chi^2$ function. 
\begin{itemize}
\item
The CKM sector favors  $\xi_q>1$. For $\xi_q\approx 1$  the hierarchy between $\eps_{q^2}$ and $\eps_{q^3}$ is not very large (see Fig.~\ref{fig:epsdist}) causing $\theta_{23}$ to turn out somewhat large, see Eq.~(\ref{eq:approxevmix}). This is improved by increasing $\xi_q$ (see Fig.~\ref{fig:epsdistb}). 
The CKM matrix can be well reproduced for  values  $2 \leq N_{q} \leq 4$.
\item
In the up sector the top quark mass prefers $\xi_q$ and $\xi_u$ of order one. 
The simultaneous fit of the ratios $m_t/m_c$ and $m_c/m_u$ hierarchies prefers slightly larger values of $\xi_q$ and/or $\xi_u$.
The up-sector masses are in general well reproduced for 
$8\lesssim N_{q} + N_{u} \lesssim 12$, depending on the values of $\xi_{q,u}$.
\item
The down sector prefers values of $\xi_q$ and/or $\xi_d$ greater than one in order to suppress the bottom mass. Given the restriction on $\xi_q$ from the top mass, $\xi_d$ typically is the larger of the two. A larger $N_d$ would help to keep $\xi_d$ smaller, but this is limited by the comparatively mild hierarchy in the down sector which prefers $5\lesssim N_q+N_d\lesssim 7$.
\end{itemize}

The most noticeable tension in the quark distributions is coming from the value of $\xi_q$ due to the opposite interests of the CKM and up sectors.\footnote{One can in principle also remove this tension by further tweaking the prior distributions for the proto-Yukawas $C_{ij}$, also allowing for values slightly larger than one.}

Taking all the previously mentioned factors into consideration one can arrive at the best fit values for the parameters by finding the minimum value of $\chi^2$. 
We consider two scenarios: Scenario A, which assumes that $N_u=N_q$ and $\xi_u=\xi_q$, naively compatible with $SU(5)$ unification, and scenario $B$, in which we set $\xi_u=1$. Both cases have just two continuous free parameters, $\xi_q$ and $\xi_d$ which we adjust to minimize $\chi^2$.
For several choices of the $N_{q,u,d}$ we present the best fit points in Table \ref{TableQuark1}.
The typical values for the $\chi^2$ are of the order of $\chi^2\sim 10$, which, for 10 degrees of freedom, corresponds to less than one $\sigma$ deviation from the mean value of the distribution.~\footnote{We stress that this $\sigma$ is the one pertaining to the theoretical distribution. By no means are we claiming that the model predicts the physical values within one experimental $\sigma$.} 
Furthermore, in Tab.~\ref{TableQuark1} we also report on a third scenario in which we set all the parameters $\xi_{q,u,d}=1$ and instead upgrade to a type-II two Higgs doublet model (2HDM) with $\tan\beta=40$ in order to accomodate the bottom mass. A noteworthy feature is that with $\xi_{q,u,d}=1$ one in general needs more clockwork gears in order to achieve a large enough hierarchy between second and third generations. The resulting values for $\chi^2$ are again of the order of one sigma, however they do not have any free continuous parameter besides $\tan\beta$.
To put these values into perspective, we have computed the corresponding distribution for $N_{q,u,d}=0$, that is, the distribution for masses and mixings in the SM if the Yukawa couplings were taken as order one random complex numbers. This yields  $\chi^2\approx 4000$, putting the physical values at about 63 sigma away from the mean value of this distribution.  
 Including $\tan\beta$, these values "improve" to $\chi^2\approx 1600$ or 39 sigma.

\begin{figure}[!htb]%
    \centering
    \subfloat[Down sector Yukawa eigenvalues]{{\includegraphics[width=7cm]{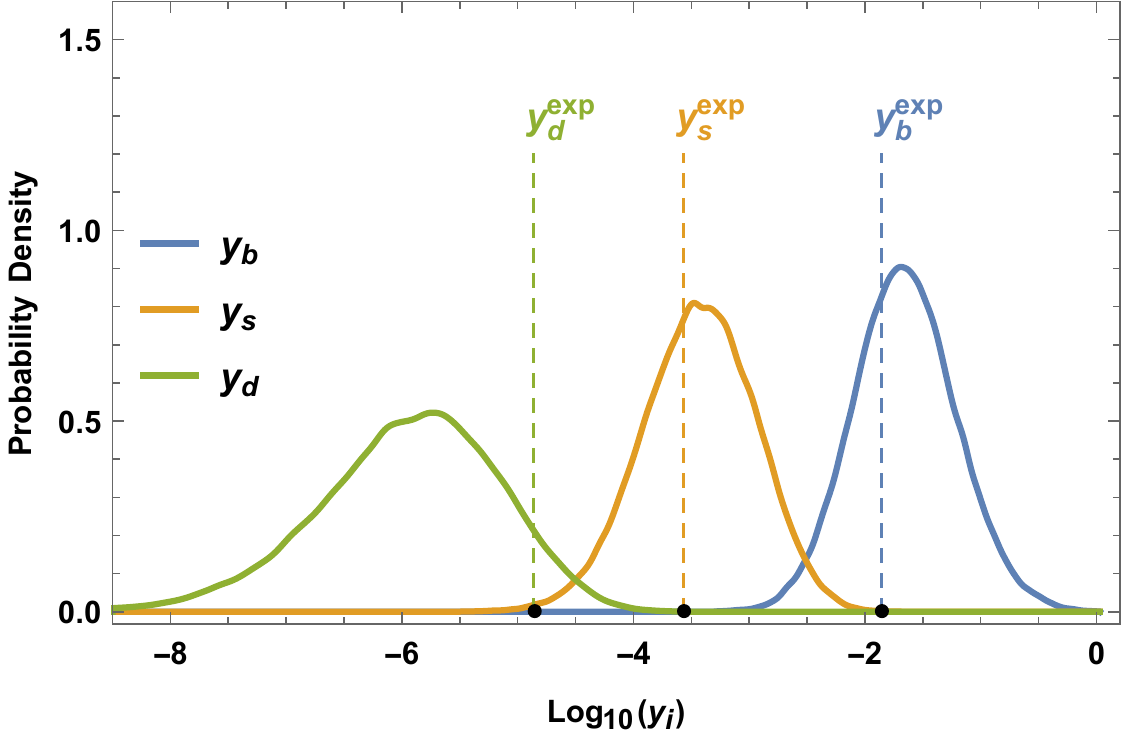}}}%
    \quad
    \subfloat[htc][Up sectorYukawa eigenvalues]{{\includegraphics[width=7cm]{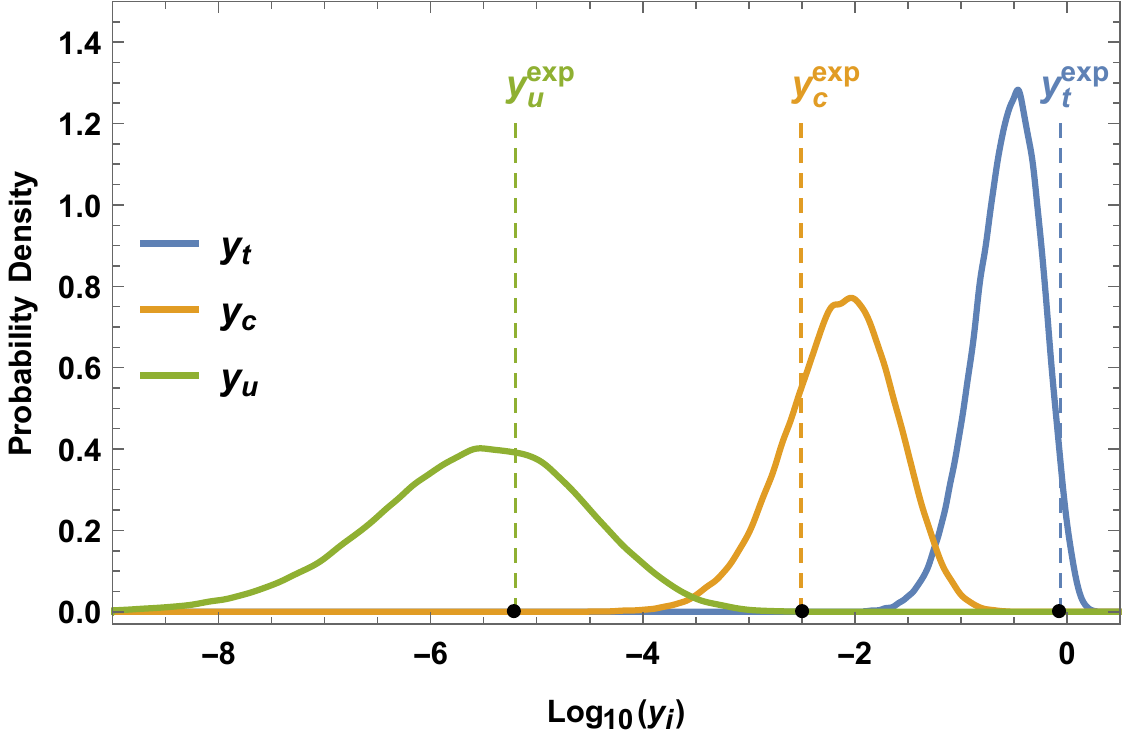}}}%
    \caption{Quark Yukawa eigenvalues distributions for $N_{q}=5$, $N_{u}=7$, $N_{d}=2$, $\xi_{q}=1.8$, $\xi_{u}=1$ and $\xi_{d}=10$. }%
    \label{fig:SimQuark1}%
\end{figure}

\begin{figure}[!htb]%
    \centering
    \subfloat[htc][CKM mixing angles]{{\includegraphics[width=7cm]{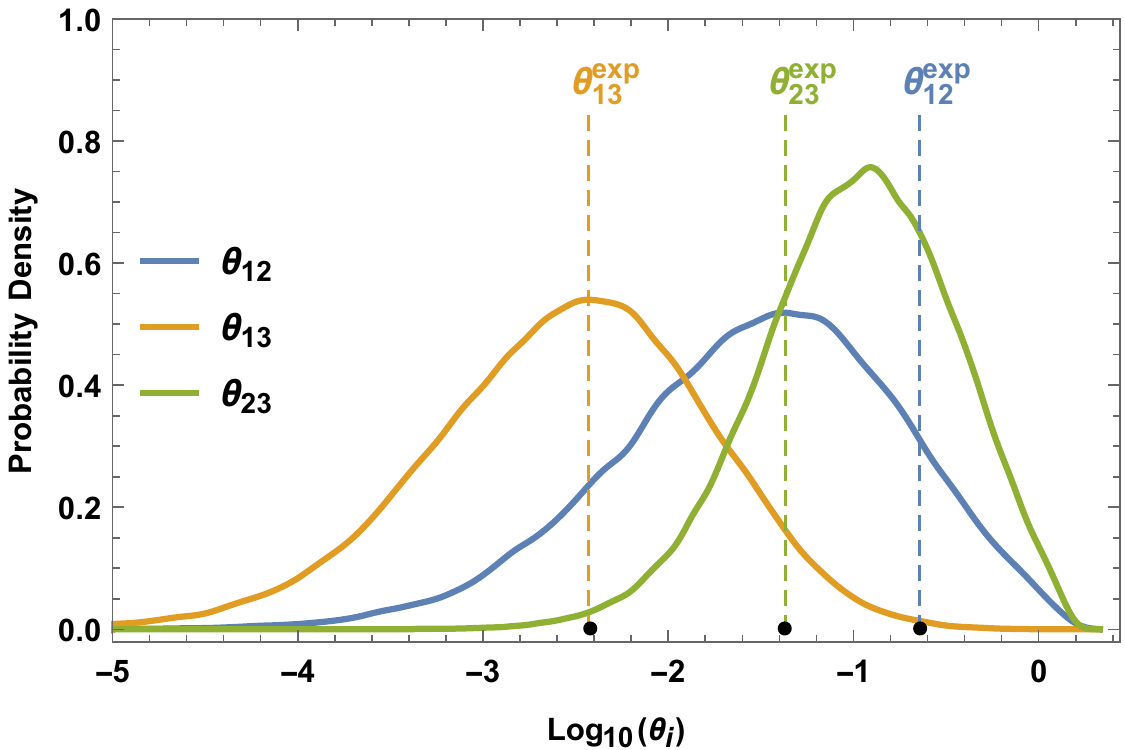}}}%
    \quad
    \subfloat[htc][Jarlskog invariant]{{\includegraphics[width=7cm]{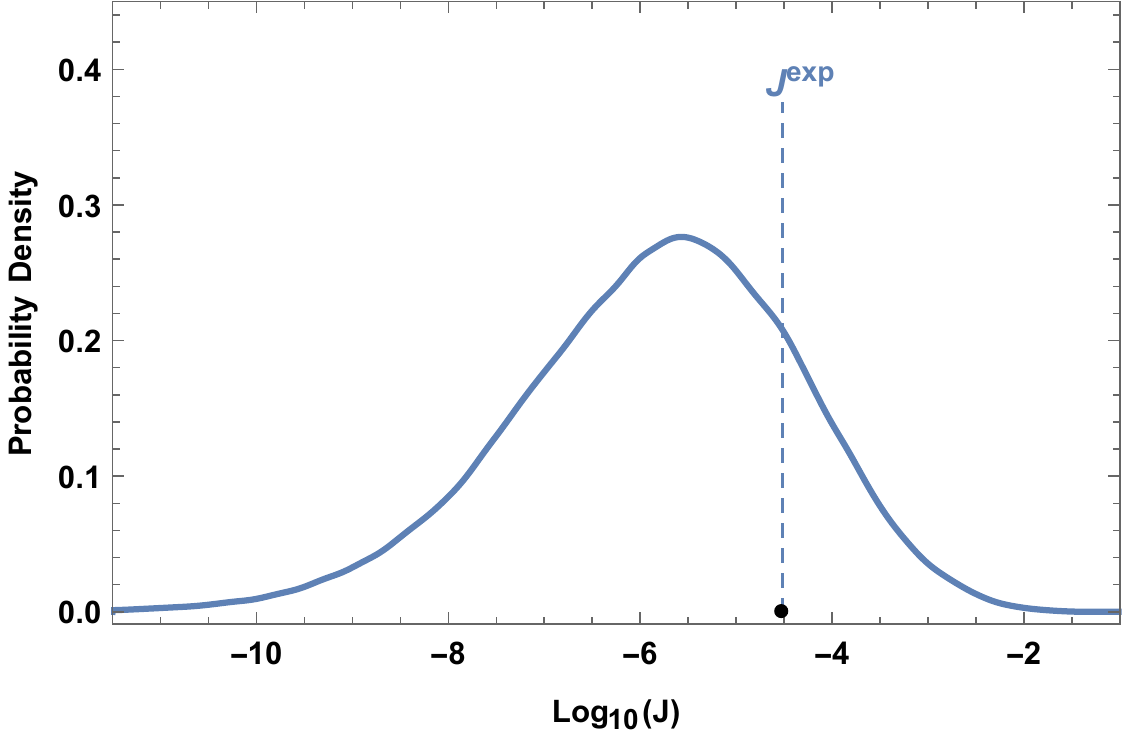}}}%
    \caption{Quark mixing distributions for $N_{q}=5$, $N_{u}=7$, $N_{d}=2$, $\xi_{q}=1.8$, $\xi_{u}=1$ and $\xi_{d}=10$.}%
    \label{fig:SimQuark2}%
\end{figure}

 The marginalized distributions for a representative case can be found in Fig.~\ref{fig:SimQuark1} and Fig.~\ref{fig:SimQuark2}.

\begin{table}[!htb]
\centering
\begin{tabular}{cccccccc|cc}
	\toprule
	   &  \multicolumn{3}{c}{Scenario A}	& \multicolumn{3}{c}{Scenario B}& 2HDM &SM & 2HDM \\
	\midrule
	 $N_{q}$ 	&	5	&	4	&	4	&	5	&	$3$	&	4 	&	10	&0	&0\\
	 $N_{u}$ 	&	5	&	4	&	4	&	7	&	$8$	&	8	&	10	&0	&0\\
	 $N_{d}$ 	&	2	&	2	&	3	&	2	&	$2$	&	3	&	4	&0	&0\\
	 \midrule
	 $\xi_{q}$	&	1.5	&	1.8	&	1.8	&	1.8	&	$3$	&	2  	&	1	&-	&-\\
	 $\xi_{u}$	&	1.5	&	1.8	&	1.8	&	1	&	$1$	&	1 	&	1	&-	&-\\
	 $\xi_{d}$	&	13	&	12	&	5.5	&	10	&	$9$	&	5	&	1	&-	&-\\
	\midrule
	$\chi_{\rm quark}^{2}$	&	9.4	&	10.2	&	11.7	&	9.3	&	10.3	&	10.5	&	10.5	&4000&1600\\
	$\sigma_{\rm quark}$  &    0.7	&	0.8	&	1.0	&	0.7	&	0.8	&	0.8	&	0.8	&63	&39\\
	\bottomrule
	\end{tabular}
\caption{Values of $\chi^2$ for several choices $N_{q,u,d}$ and  $\xi_{q,u,d}$. 
The ten degrees of freedom involved in the computation of $\chi^{2}$ are the six quark masses, the three mixing angles and the Jarlskog invariant $J$. For the 2HDM we set $\tan\beta=40$.}
\label{TableQuark1}
\end{table}

\subsection{Leptons}

Moving to the lepton sector, we have the two discrete ($N_{\ell}$, $N_{e}$), and three continuous parameters
 $(\xi_{\ell}$, $\xi_{e}$, $m_{\nu_R}$). For the neutrino sector, we consider normal mass ordering only.
To good approximation, the charged lepton masses only depend on $\xi_e$, $\xi_\ell$ , $N_e$ , and $N_\ell$, while the mixings mainly depend on $N_\ell$ (and weakly on $\xi_\ell$) and the neutrino masses only depend on $\xi_\ell$, $N_\ell$ and $m_{\nu_R}$.

The following features determine the main behaviour of the $\chi^2$ function.
\begin{itemize}
\item
The PMNS matrix requires a low value of $N_\ell$  in order to avoid a too small $\theta_{13}$:  $N_\ell=1,2$ work best, with $N_\ell>  3$ noticeably deteriorating the fit. The dependence on $\xi_\ell$ is  weak, the main effect of larger $\xi_\ell$ is a mild suppression of  $\theta_{23}$.
\item
The charged leptons require a somewhat large value of $\xi_e$ and/or $\xi_\ell$ in order to suppress the $\tau$ mass.
\item
For the neutrino sector, one needs to avoid large hierarchies between the second and third generations ($m_{\nu^3}/m_{\nu^2}\lesssim 6$ for normal ordering), which points towards $\xi_\ell\lesssim 1$. However, for small $N_\ell$ (preferred by PMNS, see above), even a large $\xi_\ell$ is in principle possible. Since this decreases the neutrino masses, it requires a smaller seesaw scale in order to achieve a large enough $m_{\nu^3}$.
\end{itemize}

\begin{figure}[!htb]%
    \centering
    \subfloat[Charged leptons sector Yukawa eigenvalues]{{\includegraphics[width=7cm]{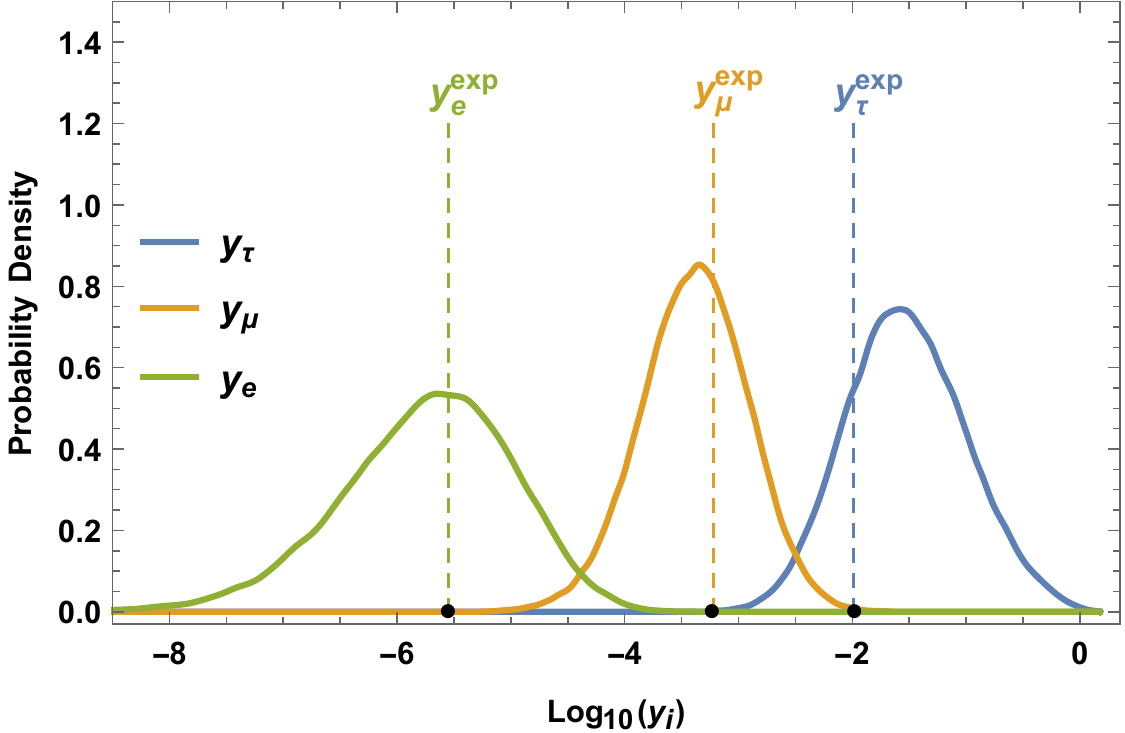}}}%
    \quad
    \subfloat[htc][Neutrino squared masses]{{\includegraphics[width=7cm]{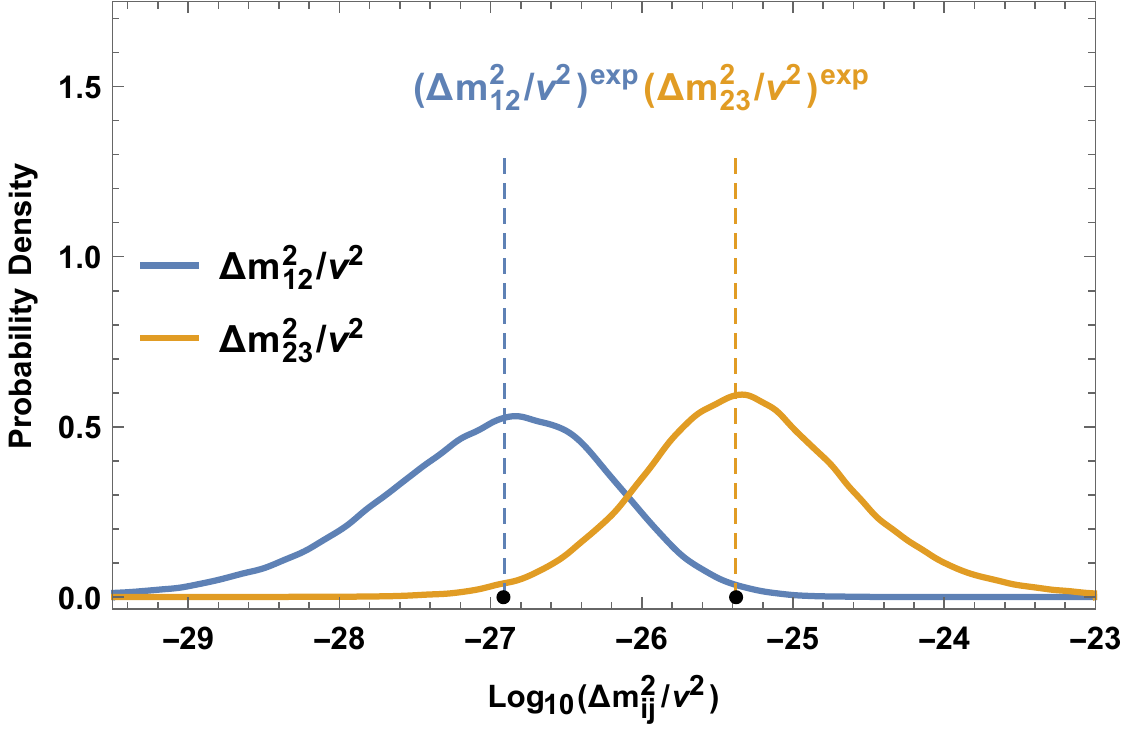}}}%
    \caption{Lepton masses distributions for $N_{\ell}=2$, $N_{e}=5$, $\xi_{\ell}=1$ and $\xi_{e}=4$. Both neutrinos and charged leptons masses are very well reproduced.}%
    \label{fig:SimLepton1}%
\end{figure}

\begin{figure}[!htb]%
    \centering
    \subfloat[htc][PMNS mixing angles]{{\includegraphics[width=7cm]{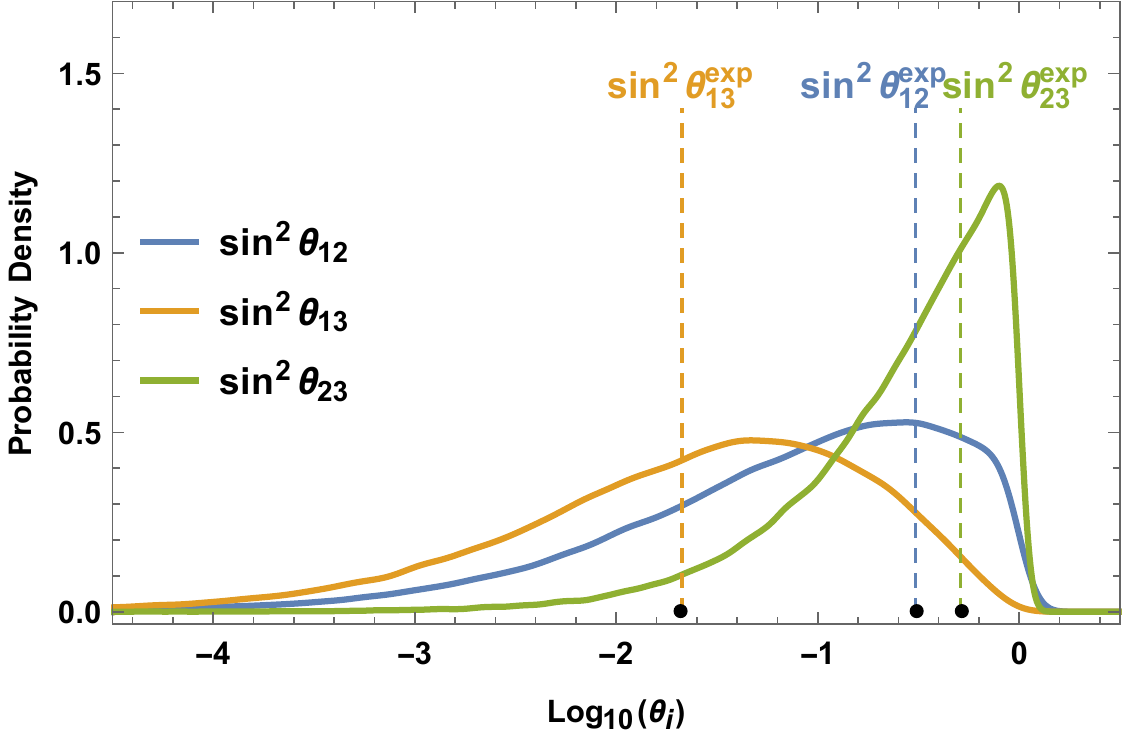}}}%
    \quad
    \subfloat[htc][Rephasing invariant]{{\includegraphics[width=7cm]{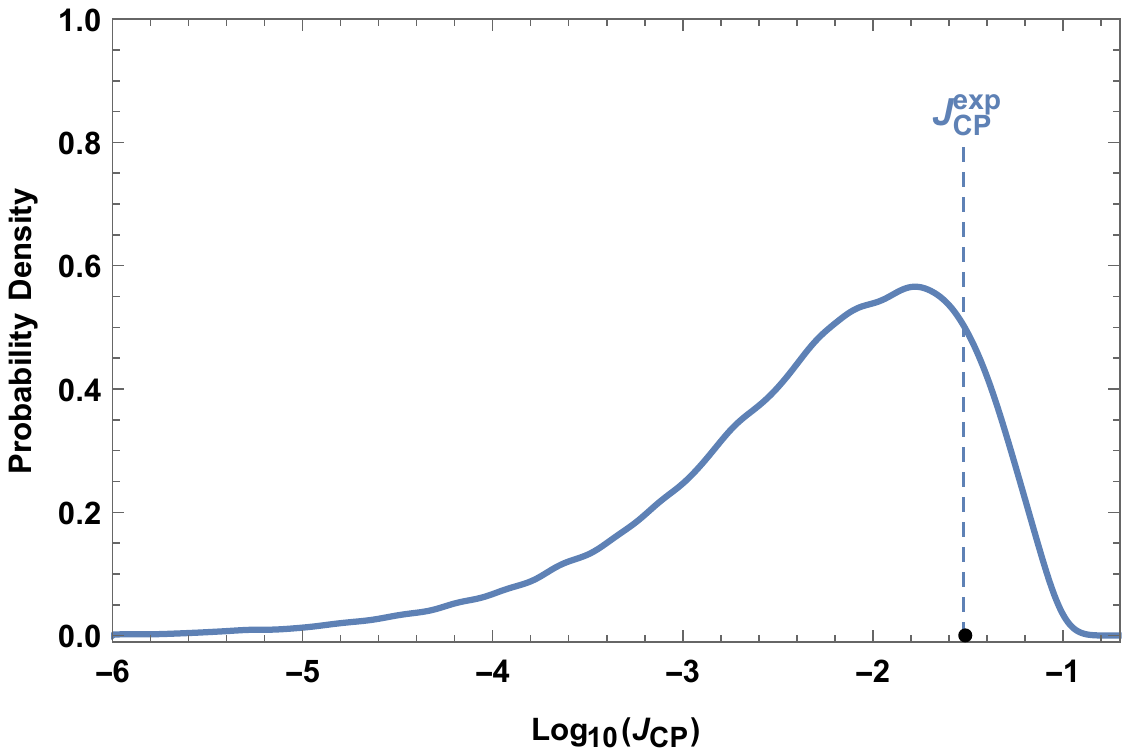}}}%
    \caption{Lepton mixing distributions for $N_{\ell}=2$, $N_{e}=5$, $\xi_{\ell}=1$ and $\xi_{e}=4$.}%
    \label{fig:SimLepton2}%
\end{figure}

The lepton sector is thus very easy to accommodate. 
Our results for the leptons are summarized in Tab.~\ref{tab:lep}, and some  distributions can be found in Figs.~\ref{fig:SimLepton1} and \ref{fig:SimLepton2}. 
The $SU(5)$ compatible scenario A has all parameters fixed from the quark sector, $N_e=N_q$, $N_\ell=N_d$, $\xi_e=\xi_q$, and $\xi_\ell=\xi_d$. This leaves as the only free parameter  the seesaw scale, which we use to  the 
 fit the neutrino masses. As the neutrino masses are suppressed by  the relatively large values of $\xi_\ell$, the seesaw scale is comparatively small,   $\sim 5\cdot  10^{11}$ GeV. 
 For the scenario $B$, we fix $\xi_{\ell}=1$ and vary only $m_{\nu_R}$  and $\xi_e$.
Notice that the first three cases of scenario B are still compatible with $SU(5)$ as far as the  number of gears go, and only show $SU(5)$ violation in the $\xi_{e,\ell}$. For the 2HDM scenario, we take the same $SU(5)$ compatible choices from the quark sector, 
leaving again only one free parameter, the seesaw scale. 

Again, Tab.~\ref{tab:lep} contains the "random SM" and its  2HDM cousin for comparison. 
It is worth noticing that in both cases almost the entire contribution to the $\chi^2$ comes from the charged lepton masses, with the neutrino sector adding very little. This simply shows that neutrino anarchy \cite{Hall:1999sn,Haba:2000be} is working well.

\begin{table}[!htb]
\centering
\begin{tabular}{c ccc  ccccc c|cc}
	\toprule
	 	 		&  \multicolumn{3}{c}{Scenario A}&  \multicolumn{5}{c}{Scenario B}& 2HDM&SM&2HDM \\
	\midrule
	 $N_{\ell}$ 	&	2	&	2	&3		&	\ 2	&	2	&3	&1	&1		& 4		&0		&0\\
	 $N_{e}$ 		&	5	&	4	&4		&	\ 5	&	4	&4	&5	&6		& 10		&0		&0\\
	 \midrule
	 $\xi_{\ell}$	&	13	&	12	&5.5		&	\ 1	&	1	&1	&1	&1		& 1		&-		&-\\
	 $\xi_{e}$		&	1.5	&	1.8	&1.8		&	\ 4	&	6	&6	&4	&3		& 1		&-		&-\\
	 $m_{\nu_R}$ 	
	 			&\multicolumn{3}{c}{$5\cdot 10^{11}$ GeV}	&\multicolumn{5}{c}{$10^{15}$ GeV}&$10^{15}$ GeV	&\multicolumn{2}{c}{$10^{15}$ GeV}	\\
	\midrule
	$\chi^{2}_{\rm lepton}$		&	5.2	& 	5.5	&7.3		&\	3.5	& 	3.2	&3.8		&4.1	&5.7		& 5.3		&3000	&570\\
	$\sigma_{\rm lepton}$ 		&	0.2	&	0.3	& 0.5  	&\	0.1	&	0.1	&0.1 	&0.1	&0.3		& 0.2	&54		&23\\
	\bottomrule
\end{tabular}
\caption{Values of $\chi^2$ for several choices of $N_{\ell,e}$ and $\xi_{\ell,e}$, and the seesaw scale $m_{\nu_R}$. 
The nine degrees of freedom involved in the computation of $\chi^{2}$ are the three charged lepton  masses, the three PMNS mixing angles and the Jarlskog invariant $J$. For the 2HDM we set $\tan\beta=40$.}
\label{tab:lep}
\end{table}

\section{Dimension-six operators}
\label{sec:dim6}

In this section we compute the dimension-six operators relevant for the leading flavor changing neutral currents (FCNC) effects of the model.
Writing the effective theory as
\be
\mathcal L_6= \sum_I C_I\mathcal O_I\,,
\ee
our task is to compute the Wilson coefficients $C_I$, 
where we will be adopting the Warsaw basis \cite{Grzadkowski:2010es}.

\label{sec:ops}

\subsection{Tree level}

 All tree level flavor changing effects come from the dimension-six effective Lagrangian (in canonical normalization)
\be
\mathcal L_{\rm 6}=-i\sum_{\psi} \bar \psi (\E_\psi A_{\psi^0}\E_\psi)\Dsl^3  \psi\,.
\label{eq:L6}
\ee
Here, $A_\psi$ is the order $p^2$ term in the expansion of the form factors Eq.~(\ref{eq:ffexp}), that is given recursively in terms of
\be
A_{\psi^{\ell-1}}=(Q^\psi_\ell)^\dagger (A_{\psi^{\ell}}+Z_{\psi^\ell}|M^\psi_{\ell}|^{-2}Z_{\psi^\ell})Q^\psi_\ell\,,\qquad A_{\psi^N}=0\,.
\ee
One can solve this explicitly to get 
\be
A_{\psi^0}=\sum_{\ell=1}^{{N_\psi}} (Q^\psi_{\ell}\dots Q^\psi_1)^\dagger Z_{\psi^{\ell}}|M_{\ell}^\psi|^{-2}Z_{\psi^{\ell}}(Q^\psi_{\ell}\dots Q^\psi_1)\,,
\ee
where $Z_{\psi^\ell}$ was given explicitly in Eq.~(\ref{eq:Z}).

\begin{table}
\begin{center}
\begin{tabular}
{cc c }
\toprule
Operator 	& Wilson coefficient (exact) 	& Wilson coefficient (approx.)\\
\midrule
$\mathcal (O_{eH})^{ij}=|H|^2\bar\ell^i He^j$	&	$C_{eH}=\frac{1}{2}\left(Y_eA_eY_e^\dagger Y_e+Y_eY_e^\dagger A_\ell Y_e\right)$	& $( C_{eH})_{ij}\sim \eps_{\ell^i}\eps_{e^j}c_e^3\, \mc_{e,\ell}^{-2}$ \\
$\mathcal (O_{uH})^{ij}=|H|^2\bar q^i \tilde Hu^j $	&	$C_{uH}=\frac{1}{2}\left(Y_uA_uY_u^\dagger Y_u+Y_uY_u^\dagger A_qY_u\right)$	& $( C_{uH})_{ij}\sim \eps_{q^i}\eps_{u^j}c_u^3\mc_{u,q}^{-2} $ \\
$\mathcal (O_{dH})^{ij}=|H|^2\bar q^i Hd^j $	&	$C_{dH}=\frac{1}{2}\bigl(Y_dA_dY_d^\dagger Y_d+Y_dY_d^\dagger A_q Y_d\bigr)$	& $( C_{dH})_{ij}\sim \eps_{q^i}\eps_{d^j}c_d^3\mc_{d,q}^{-2} $ 
\\
\bottomrule
\end{tabular}
\end{center}
\caption{Yukawa corrections and their Wilson coefficients.}
\label{tab:yuk}
\end{table}%

We prefer to translate the operators in $\mathcal L_6$ to equivalent operators with fewer derivatives \cite{Grzadkowski:2010es}. Using integration by parts and the equations of motion we find
only two type of operators. The first ones are corrections to Yukawa couplings  given in Tab.~\ref{tab:yuk},
where we have defined
\be
A_\psi\equiv \E_\psi A_{\psi^0}\E_\psi\,.
\ee

The others  are current-current interactions given in Tab.~\ref{tab:cc},
\begin{table}
\begin{center}
\begin{tabular}
{ccc}
\toprule
Operator 	& Wilson coefficient (exact) 	& Wilson coefficient (approx.)\\
\midrule
$(\mathcal O_{He})^{ij}=(J^\mu_H)(\bar e^i\gamma_\mu  e^j) $	&	$C_{He}=Y_e^\dagger A_\ell Y_e$	& $( C_{He})_{ij}\sim \eps_{e^i}\eps_{e^j}c_e^2 \mc_\ell^{-2}$ \\
$(\mathcal O_{Hu})^{ij}=(J^\mu_H)(\bar u^i\gamma_\mu  u^j) $	&	$C_{Hu}=-Y_u^\dagger A_q Y_u$	& $( C_{Hu})_{ij}\sim \eps_{u^i}\eps_{u^j}c_u^2 \mc_q^{-2} $ \\
$(\mathcal O_{Hd})^{ij}=(J^\mu_H)(\bar d^i\gamma_\mu  d^j)  $	&	$C_{Hd}= Y_d^\dagger A_q Y_d$	& $( C_{Hd})_{ij}\sim \eps_{d^i}\eps_{d^j}c_d^2 \mc_q^{-2}$ \\
\midrule
$(\mathcal O^{(1)}_{H\ell})^{ij}=(J^\mu_H)(\bar \ell^{i}\gamma_\mu  \ell^j)$	& 	$C_{H\ell}^{(1)}=-\frac{1}{2}Y_eA_eY_e^\dagger$	
		& $( C^{(1)}_{H\ell}+ C^{(3)}_{H\ell})_{ij}\sim \eps_{\ell^i}\eps_{\ell^j}c_e^2\mc_e^{-2}$\\
$(\mathcal O^{(3)}_{H\ell})^{ij}=(\vec J^\mu_H)(\bar \ell^{i}\,\vec\sigma\,\gamma_\mu  \ell^j)$	&$C_{H\ell}^{(3)}=-\frac{1}{2}Y_eA_eY_e^\dagger$	
		&$(\hat C^{(1)}_{H\ell}-\hat C^{(3)}_{H\ell})_{ij}=0$\\
$(\mathcal O^{(1)}_{Hq})^{ij}=(J^\mu_H)(\bar q^{i}\gamma_\mu  q^j) $	&	 $C_{Hq}^{(1)}=-\frac{1}{2}(Y_dA_dY_d^\dagger-Y_uA_uY_u^\dagger)   $
		&$( C^{(1)}_{Hq}+ C^{(3)}_{Hq})_{ij}\sim \eps_{q^i}\eps_{q^j}c_d^2\mc_d^{-2}$\\
$(\mathcal O^{(3)}_{Hq})^{ij}=(\vec J^\mu_H)(\bar q^{i}\,\vec\sigma\,\gamma_\mu  q^j) $	& $C_{Hq}^{(3)}=-\frac{1}{2}(Y_dA_dY_d^\dagger+Y_uA_uY_u^\dagger)   $
		&$( C^{(1)}_{Hq}- C^{(3)}_{Hq})_{ij}\sim \eps_{q^i}\eps_{q^j}c_u^2\mc_u^{-2}$	\\
\midrule
$(\mathcal O_{Hud})^{ij}=(\widetilde J^H_\mu)(\bar u^i\gamma^\mu d^j)$	&$C_{Hud}=Y_u^\dagger A_qY_d$	&	$( C_{Hud})_{ij}\sim \eps_{u^i}\eps_{d^j}c_uc_d\mc_{q}^{-2}$
\\
\bottomrule
\end{tabular}
\end{center}
\caption{Current-current operators and their Wilson coefficients.}
\label{tab:cc}
\end{table}%
where we defined the weak Higgs currents
\bea
J^H_\mu&\equiv&\frac{i}{2}(H^\dagger D_\mu H- D_\mu H^\dagger  H)\,,\\
\vec J^H_\mu&\equiv&\frac{i}{2}(H^\dagger \,\vec \sigma \,D_\mu H- D_\mu H^\dagger  \,\vec \sigma \, H)\,,\\
\widetilde J^H_\mu&\equiv&i\,\tilde H^\dagger D_\mu H\,.
\eea

\begin{figure}[!t]%
    \centering
    {\includegraphics[width=4cm]{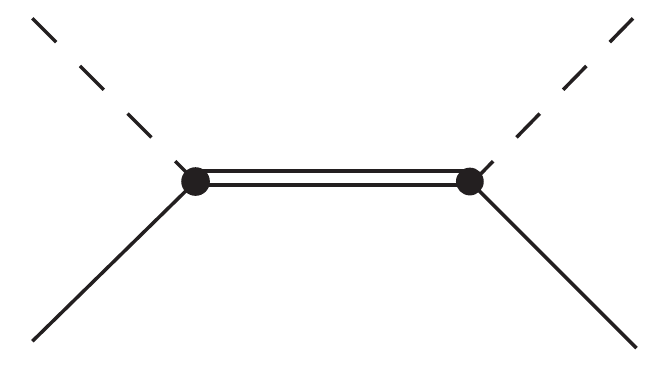}}
    \caption{Tree level contribution to the dimension-six EFT. The double line denotes a mass eigenstate before EWSB.}%
    \label{fig:tree}%
\end{figure}

In order to discuss FCNCs, it is essential to understand the flavor structure of these couplings.
Even though the expressions for the Wilson coefficients in the above operators are very explicit, they are not very illuminating for this purpose.
In App.~\ref{sec:mass} we recompute the Wilson coefficients in the mass basis, according to the diagram in Fig.~\ref{fig:tree} and show that replacing the exact mass eigenvalues by the CW scale(s) $\mc_\psi$ amounts to the substitution
\bea
\E_{\psi}A_{\psi}\E_{\psi}\to  \frac{ \one -\E_\psi^2}{\mc_\psi^2}\,,
\eea
which are anarchic matrices of order-one numbers times $\mc_\psi ^{-2}$. The quantity $\mc_\psi$ is the typical mass scale of the fermionic resonances, an explicit definition is given in Eq.~(\ref{eq:defmc}).
 In reality the mass eigenvalues lie in a band around $ \mc_\psi$ which will lead to somewhat broad distributions of allowed gear masses.
Moreover, the Yukawa couplings satisfy
\be
Y_{u,d}=c_{u,d}\, \E_q\, ({\rm order\ one\ matrix})\,  \E_{u,d}\,,
\ee
where we introduced the quantities $c_{u,d}$ which parametrize the typical size of the elements of $C_{u,d}$. Depending on the scenario, we may set $c_{u,d}=1$ or $c_u=\sin\beta $, $c_d=\cos\beta$, or incorporate any other additional parametric dependence the model may predict for the $C_{u,d}$. A similar parameter $c_e$ can be introduced in the lepton sector.

Making use of this fact, the flavor structure is now evident. 
We have, for instance
\be
C_{Hd}\equiv Y_d^\dagger A_qY_d=\left(\frac{c_d^2}{\mc_q^2}\right)\E_d\, ({\rm order\ one\ matrix})\,  \E_d\,.
\ee
The fact that the flavor violating Wilson coefficient has the same common hierarchical factor $\E_d$ as the Yukawa coupling Eq.~(\ref{eq:Yuk}) implies that the rotations that diagonalize the Yukawa couplings also approximately diagonalize the Wilson coefficients \cite{vonGersdorff:2017iym}.

By a suitable unitary (and gauge-invariant) transformation, one can obtain a basis in which the $\E_{\psi}$ are diagonal.
Let us denote the three eigenvalues by $(\eps_{\psi^i})$, with the convention that $\eps_{\psi^1}\leq \eps_{\psi^2}\leq \eps_{\psi^3}$. Notice that the eigenvalues are less than one,  $\eps_{\psi^i}\leq 1$, and are typically hierarchical.

In this basis, we can give rough estimates by omitting order-one coefficients. For instance, the Yukawa couplings themselves behave as
\be
(Y_{d})_{ij}\sim c_{d}\eps_{q^i}\eps_{{d}^j} \,,
\ee
etc.,
while the Wilson coefficient above reads
\be
(C_{Hd})_{ij}\sim 
\eps_{d^i}\eps_{d^j}
\left(\frac{c_d^2}{\mc_q^2}\right)\,.
\label{eq:Wilson}
\ee
Diagonalizing the Yukawa couplings leaves the form of  Eq.~(\ref{eq:Wilson}) unchanged.~\footnote{ More precisely it only modifies the $\mathcal O(1)$ coefficients. For the doublets $q$ and $\ell$, this change of the coefficients is different for the two components.} Therefore, this form of displaying the Wilson coefficients makes the flavor alignment very explicit, as off-diagonal entries always have at least one suppressed $\epsilon$ factor.
These parametric estimates for  all the nonzero Wilson coefficients are also given in Tab.~\ref{tab:yuk} and Tab.~\ref{tab:cc}.

\subsection{One-loop}

We now list the relevant fermionic operators that are not already generated at the tree level. 
Ignoring purely bosonic operators   not relevant for flavor observables, we focus on operators that contain either two or four fermion fields. 
The only four-fermion operators that violate flavor are generated by the diagrams in Fig.~\ref{fig:4f}. These operators are considered in Sec.~\ref{sec:4f}.
If we disregard operators that are already generated at the tree level, we can furthermore restrict the two-fermion operators to ones with at most one Higgs field.
There are two classes of operators, which schematically are of the form
$\psi^2D^3$ (two fermions, three covariant derivatives) and $ H\psi^2D^2$ (two fermions, two covariant derivatives, and one Higgs), where we count a field strength as $D^2$.
The relevant topologies are shown in Fig.~\ref{fig:dip}. 
After reduction to the Warsaw basis we are left only with  dipole type operators, operators already present at the tree level, as well as $\Delta F=1$ four-fermion operators.

Notice that in both cases only Higgs loops but not gauge loops contribute.

\subsubsection{Four-fermion operators}
\label{sec:4f}

\begin{figure}[!htb]%
    \centering
    {\includegraphics[width=4cm]{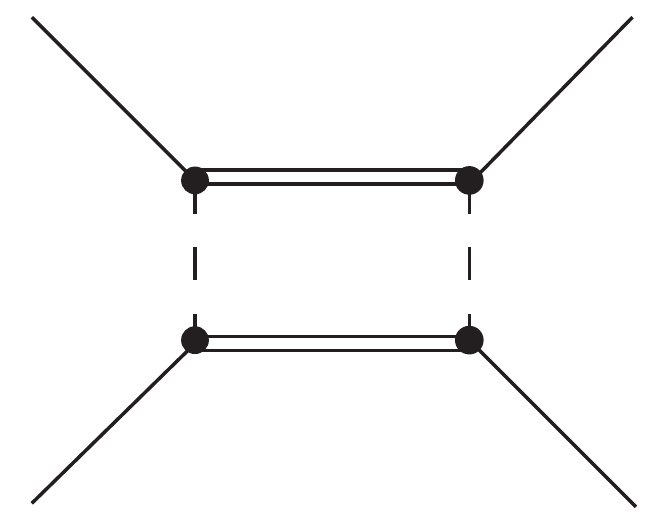}}
    \caption{Diagram contributing to four-fermion operators. The double line denotes a mass eigenstate before EWSB.}%
    \label{fig:4f}%
\end{figure}

Dimension-6 operators with four fermions  will appear at the one loop level only, and involve two virtual Higgs exchanges, see diagrams in Fig.~\ref{fig:4f}.

For convenience, we define the function
\be
M^{-2}_{ab}=\frac{1}{16\pi^2}\, \frac{\log\left(\frac{\mc_a}{\mc_b}\right)}{\mc_a^2-\mc_b^2}\,,
\ee
where $a,b=q,\ell,u,d,e$, which parametrizes the diagram with exchange of $a$ and $b$ type clockwork fermions.
These decouple with the heavier of the two scales $\mc_a$ and $\mc_b$.

In terms of the Warsaw basis, one generates all vector-current squared  operators except $\mathcal O_{qu}^{(8)}$ and $\mathcal O_{qd}^{(8)}$ and none of the scalar-current squared operators. 
Here, for simplicity, we define the alternative operator $\widetilde{\mathcal  O}_{ud}$ which is used instead of the operator $\mathcal O_{ud}^{(8)}$.

The four-fermion operators and their Wilson coefficients are given in Tab.~\ref{tab:4f}. 

\begin{table}
\begin{center}
\begin{tabular}
{cc}
\toprule
Operator 	& Wilson coefficient (approximate)\\
\midrule
$(\mathcal O_{\ell\ell})^{ijkl}=(\bar\ell^i\gamma^\mu\ell^j)(\bar\ell^k\gamma_\mu\ell^l)$	&
	$(C_{\ell\ell})_{ijkl}		\sim \eps_{\ell^i}\eps_{\ell^j}\eps_{\ell^k}\eps_{\ell^l} \, \frac{1}{2}   \,c_\ell^4   M_{ee}^{-2}$\\
$(\mathcal O_{qq}^{(1)})^{ijkl}=(\bar q^i\gamma^\mu q^j)(\bar q^k\gamma_\mu q^l)$	&	
	$(C^{(1)}_{qq})_{ijkl}		\sim \eps_{q^i}\eps_{q^j}\eps_{q^k}\eps_{q^l}\,  \frac{1}{4}( c_u^4 M_{uu}^{-2}+c_d^4M_{dd}^{-2}+2c_u^2c_d^2M_{ud}^{-2})$\\
$(\mathcal O_{qq}^{(3)})^{ijkl}=(\bar q^i\gamma^\mu \vec\sigma q^j)(\bar q^k\gamma_\mu \vec\sigma q^l)$	&
	$(C^{(3)}_{qq})_{ijkl}		\sim \eps_{q^i}\eps_{q^j}\eps_{q^k}\eps_{q^l}\,  \frac{1}{4}( c_u^4 M_{uu}^{-2}+c_d^4M_{dd}^{-2}+2c_u^2c_d^2M_{ud}^{-2})$\\
$(\mathcal O_{\ell q}^{(1)})^{ijkl}=(\bar \ell^i\gamma^\mu \ell^j)(\bar q^k\gamma_\mu q^l)$	&
	$(C^{(1)}_{\ell q})_{ijkl}	\sim \eps_{\ell^i}\eps_{\ell^j}\eps_{q^k}\eps_{q^l}\,  \frac{1}{4}(c_u^2c_e^2M_{ue}^{-2}+c_d^2c_e^2M_{de}^{-2})$\\
$(\mathcal O_{\ell q}^{(3)})^{ijkl}=(\bar \ell^i\gamma^\mu \vec\sigma \ell^j)(\bar q^k\gamma_\mu \vec\sigma q^l)$	&
	$(C^{(3)}_{\ell q})_{ijkl}	\sim \eps_{\ell^i}\eps_{\ell^j}\eps_{q^k}\eps_{q^l}\,  \frac{1}{4}(c_u^2c_e^2M_{ue}^{-2}+c_d^2c_e^2M_{de}^{-2})$\\
		\midrule
$(\mathcal O_{ee})^{ijkl}=(\bar e^i\gamma^\mu e^j)(\bar e^k\gamma_\mu e^l)$	&
	$(C_{ee})_{ijkl}		\sim \eps_{e^i}\eps_{e^j}\eps_{e^k}\eps_{e^l}\,c_e^4  M_{\ell\ell}^{-2}$\\		
$(\mathcal O_{uu})^{ijkl}=(\bar u^i\gamma^\mu u^j)(\bar u^k\gamma_\mu u^l)$	&
	$(C_{uu})_{ijkl}		\sim \eps_{u^i}\eps_{u^j}\eps_{u^k}\eps_{u^l}\, c_u^4  M_{qq}^{-2}$\\		
$(\mathcal O_{dd})^{ijkl}=(\bar d^i\gamma^\mu d^j)(\bar d^k\gamma_\mu d^l)$	&
	$(C_{dd})_{ijkl}		\sim \eps_{d^i}\eps_{d^j}\eps_{d^k}\eps_{d^l}\,  c_d^4 M_{qq}^{-2}$\\		
$(\mathcal O_{eu})^{ijkl}=(\bar e^i\gamma^\mu e^j)(\bar u^k\gamma_\mu u^l)$	&
	$(C_{eu})_{ijkl}		\sim \eps_{e^i}\eps_{e^j}\eps_{u^k}\eps_{u^l}\,  c_e^2c_u^2 M_{\ell q}^{-2}$\\		
$(\mathcal O_{ed})^{ijkl}=(\bar e^i\gamma^\mu e^j)(\bar d^k\gamma_\mu d^l)$	&
	$(C_{ed})_{ijkl}		\sim \eps_{e^i}\eps_{e^j}\eps_{d^k}\eps_{d^l}\,  c_e^2c_d^2 M_{\ell q}^{-2}$\\		
$(\mathcal O_{ud})^{ijkl}=(\bar u^i\gamma^\mu  u^j)(\bar d^k\gamma_\mu d^l)$	&
	$(C_{ud})_{ijkl}		\sim \eps_{u^i}\eps_{u^j}\eps_{d^k}\eps_{d^l}\,  c_u^2c_d^2 M_{qq}^{-2}$\\				
$(\widetilde{\mathcal O}_{ud})^{ijkl}=(\bar u^i\gamma^\mu  d^j) 	(\bar d^k\gamma_\mu u^l)$	&
	$(\widetilde C_{ud})_{ijkl}	\sim \eps_{u^i}\eps_{d^j}\eps_{d^k}\eps_{u^l}\, c_u^2c_d^2  M_{qq}^{-2}$\\		
	\midrule	
$(\mathcal O_{\ell e})^{ijkl}=(\bar \ell^i\gamma^\mu  \ell^j)(\bar e^k\gamma_\mu e^l)$	&
	$(C_{\ell e})_{ijkl}	\sim \eps_{\ell^i}\eps_{\ell^j}\eps_{e^k}\eps_{e^l}\,   \frac{1}{2} c_e^4M_{e\ell}^{-2}$\\				
$(\mathcal O_{\ell u})^{ijkl}=(\bar \ell^i\gamma^\mu  \ell^j)(\bar u^k\gamma_\mu u^l)$	&
	$(C_{\ell u})_{ijkl}	\sim \eps_{\ell^i}\eps_{\ell^j}\eps_{u^k}\eps_{u^l}\,   \frac{1}{2} c_e^2c_u^2M_{eq}^{-2}$\\				
$(\mathcal O_{\ell d})^{ijkl}=(\bar \ell^i\gamma^\mu  \ell^j)(\bar d^k\gamma_\mu d^l)$	&
	$(C_{\ell d})_{ijkl}	\sim \eps_{\ell^i}\eps_{\ell^j}\eps_{d^k}\eps_{d^l}\,   \frac{1}{2} c_e^2c_d^2M_{eq}^{-2}$\\
$(\mathcal O_{q e})^{ijkl}=(\bar q^i\gamma^\mu  q^j)(\bar e^k\gamma_\mu e^l)$	&
	$(C_{qe})_{ijkl}	\sim \eps_{q^i}\eps_{q^j}\eps_{e^k}\eps_{e^l}\,  \frac{1}{2}  (c_u^2c_e^2M_{u\ell}^{-2}+c_d^2c_e^2M_{d\ell}^{-2})$\\								
$(\mathcal O_{q u})^{ijkl}=(\bar q^i\gamma^\mu  q^j)(\bar u^k\gamma_\mu u^l)$	&
	$(C_{qu})_{ijkl}	\sim \eps_{q^i}\eps_{q^j}\eps_{u^k}\eps_{u^l}\, \frac{1}{2}   (c_u^4M_{uq}^{-2}+c_u^2c_d^2M_{dq}^{-2})$\\								
$(\mathcal O_{q d})^{ijkl}=(\bar q^i\gamma^\mu  q^j)(\bar d^k\gamma_\mu d^l)$	&
	$(C_{qd})_{ijkl}	\sim \eps_{q^i}\eps_{q^j}\eps_{d^k}\eps_{d^l}\, \frac{1}{2}   (c_u^2c_d^2M_{uq}^{-2}+c_d^4M_{dq}^{-2})$\\								
\bottomrule
\end{tabular}
\end{center}
\caption{Four-fermion operators and their Wilson coefficients.}
\label{tab:4f}
\end{table}%

\subsubsection{Two-fermion operators}
\label{sec:dip}

\begin{figure}[htb]%
    \centering
    \subfloat[]{{\includegraphics[width=4.5cm]{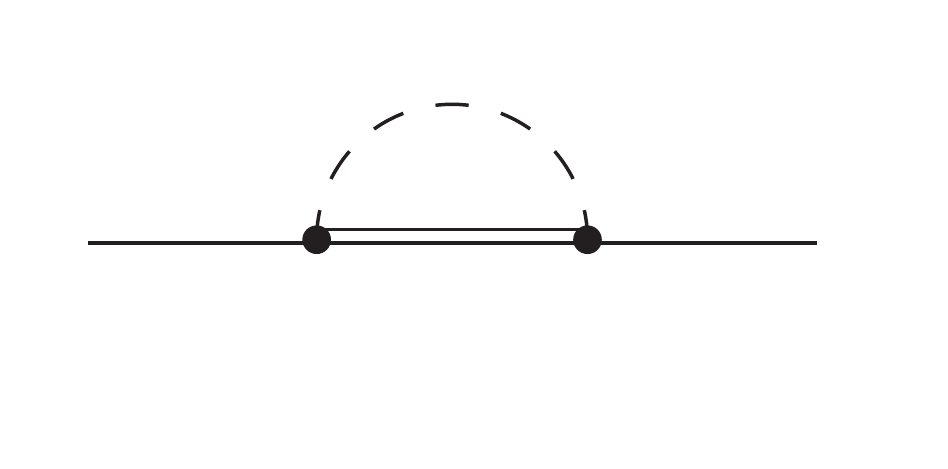}}}%
    \quad
    \subfloat[]{{\includegraphics[width=4.5cm]{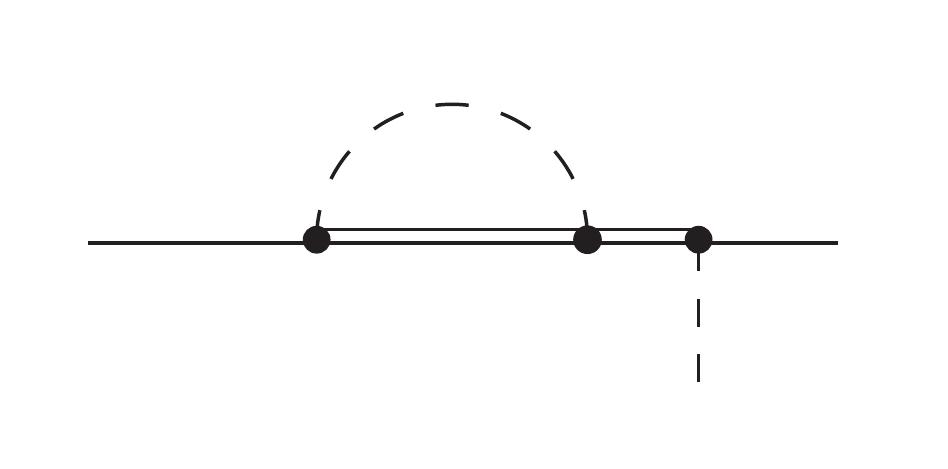}}}%
    \quad
    \subfloat[]{{\includegraphics[width=4.5cm]{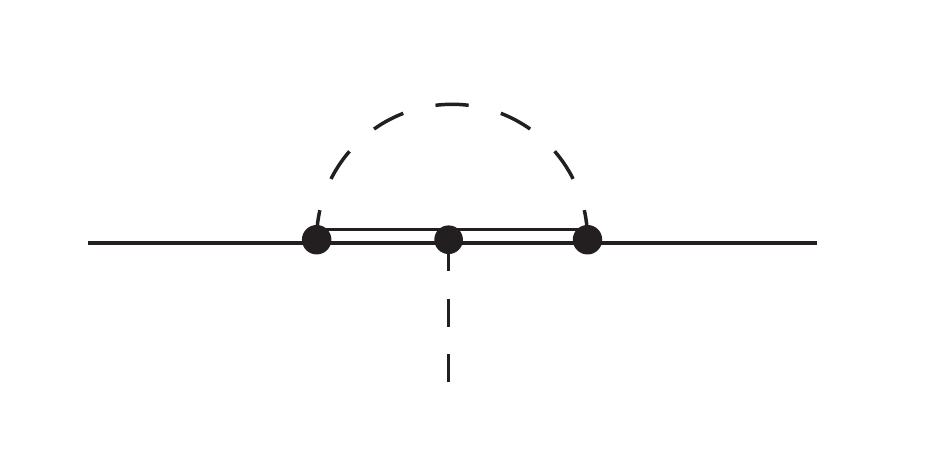}}}
    \caption{Diagrams contributing to dimension-six operators with two fermions. The double line denotes a mass eigenstate before EWSB.
    Different numbers of gauge boson lines can be inserted at any of the internal lines without affecting the flavor structure.}%
    \label{fig:dip}%
\end{figure}

\begin{table}
\begin{center}
\begin{tabular}
{cc}
\toprule
Operator 	 	& Wilson coefficient (approximate)\\
\midrule
$(\mathcal O_{eB})^{ij}=\bar \ell^i H \sigma^{\mu\nu} e^j B_{\mu\nu} $
	&$( C_{eB})_{ij}\sim \eps_{\ell^i}\eps_{e^j}\frac{g'c_e^3}{16\pi^2} \frac{1}{\mc_{e,\ell}^{2}}$	\\	
$(\mathcal O_{uB})^{ij}=\bar q^i H \sigma^{\mu\nu} u^j B_{\mu\nu} $
	&$( C_{uB})_{ij}\sim\eps_{q^i}\eps_{u^j}\frac{g'c_u}{16\pi^2}  \left( \frac{c_u^2}{\mc_{q,u}^2}+ \frac{c_d^2}{\mc_{q,d}^2} \right)$\\
$(\mathcal O_{dB})^{ij}=\bar q^i H \sigma^{\mu\nu} d^j B_{\mu\nu} $
	&$( C_{dB})_{ij}\sim\eps_{q^i}\eps_{d^j}\frac{g'c_d}{16\pi^2}  \left( \frac{c_u^2}{\mc_{q,u}^2}+ \frac{c_d^2}{\mc_{q,d}^2} \right)$\\
\midrule
$(\mathcal O_{eW})^{ij}=\bar \ell^i H \sigma^{\mu\nu} \sigma^a e^j W^a_{\mu\nu}$
	& $( C_{eW})_{ij}\sim \eps_{\ell^i}\eps_{e^j}\frac{gc_e^3}{16\pi^2} \frac{1}{\mc_{e,\ell}^{2}}$	\\			
$(\mathcal O_{uW})^{ij}=\bar q^i H \sigma^{\mu\nu} \sigma^a u^j W^a_{\mu\nu} $
	&$( C_{uW})_{ij}\sim\eps_{q^i}\eps_{u^j}\frac{gc_u}{16\pi^2}  \left( \frac{c_u^2}{\mc_{q,u}^2}+ \frac{c_d^2}{\mc_{q,d}^2} \right)$\\
$(\mathcal O_{dW})^{ij}=\bar q^i H \sigma^{\mu\nu} \sigma^a d^j W^a_{\mu\nu} $
	&$( C_{dW})_{ij}\sim\eps_{q^i}\eps_{d^j}\frac{gc_d}{16\pi^2}  \left( \frac{c_u^2}{\mc_{q,u}^2}+ \frac{c_d^2}{\mc_{q,d}^2} \right)$\\
\midrule
$(\mathcal O_{uG})^{ij}=\bar q^i H \sigma^{\mu\nu} \lambda^a u^j G^a_{\mu\nu} $
	&$( C_{uG})_{ij}\sim\eps_{q^i}\eps_{u^j}\frac{g_s c_u}{16\pi^2}  \left( \frac{c_u^2}{\mc_{q,u}^2}+ \frac{c_d^2}{\mc_{q,d}^2} \right)$\\
$(\mathcal O_{dG})^{ij}=\bar q^i H \sigma^{\mu\nu} \lambda^a d^j G^a_{\mu\nu} $
	&$( C_{dG})_{ij}\sim\eps_{q^i}\eps_{d^j}\frac{g_sc_d}{16\pi^2}  \left( \frac{c_u^2}{\mc_{q,u}^2}+ \frac{c_d^2}{\mc_{q,d}^2} \right)$\\
\bottomrule
\end{tabular}
\end{center}
\caption{Dipole operators and their Wilson coefficients. Here the scale $\mc_{a,b}$ is the smaller of the two scales $\mc_a$, $\mc_b$.}
\label{tab:dip}
\end{table}%

The complete set of dimension-six dipole operators is given in the first column of Tab.~\ref{tab:dip}.
All three type of diagrams in Fig.~\ref{fig:dip} in principle contribute to the dipole operators after reduction to the Warsaw basis.
One could evaluate these contributions exactly and subsequently perform an approximation of the kind Eq.~(\ref{eq:approxcw}). 
 However, as we  already plan to  ignore order-one coefficients anyway, it is sufficient to just know the parametric behaviour with the clockwork scales in the decoupling limit, which is easy to find without a detailed computation of the diagrams.
 When evaluating this decoupling behaviour, it is important to keep in mind that the clockwork gears only possess Yukawa couplings of the same chirality as in the SM, i.e., couplings of the kind $\bar q'_RHd'_L$ are absent. Therefore, between each pair of Yukawa vertices only the chirality preserving part of the Dirac propagator contributes.

Let us first consider the quark  operators. 
 The first diagrams produce $\bar q qD^3$ operators decoupling as $c_d^2 \mc_d^{-2}+c_u^2\mc_u^{-2}$, and $\bar u uD^3$ ($\bar ddD^3$) operators going as $c_{u}^2 \mc_q^{-2}$  ($c_{d}^2 \mc_q^{-2}$). By use of the equations of motion they pick up factors of $c_{u,d}$ and contribute to the dipole operators.
The second diagrams give $\bar q u \tilde H D^2$ operators going as $c_u(c_d^2+c_u^2)\mc_q^{-2}$ and $c_u^3\mc_u^{-2}$, and similarly for the $\bar q d  H D^2$ operators. Finally, the third diagrams give rise to $\bar q u \tilde H D^2$ operators proportional to $c_uc_d^2$ decoupling with the heavier of $\mc_q, \mc_d$ (and similarly for $u\leftrightarrow d$). This contribution is at most of the same order as the first two.
 The leading contributions to the Wilson coefficients therefore are
\be
g'^{-1}( C_{uB})_{ij}= g^{-1}( C_{uW})_{ij}= g_s^{-1}(  C_{uG})_{ij}\sim
\eps_{q^i}\eps_{u^j}\frac{c_u}{16\pi^2}  \left( \frac{c_u^2}{\mc_{q,u}^2}+ \frac{c_d^2}{\mc_{q,d}^2} \right)\,,
\ee
where $\mc_{a,b}$ should be taken as the smaller of the two scales,~\footnote{If the two scales are vastly different, the estimate should be multiplied by a factor $\log(\mc_a/\mc_b)$.}
and analogously for the down quark operators with $d\leftrightarrow u$.
The lepton sector works in a similar way. The resulting estimates for the Wilson coefficients are summarized in Tab.~\ref{tab:dip}.

\section{FCNC Constraints}

\label{sec:fcnc}

In this section we provide the leading constraints from FCNC observables. The analysis is somewhat similar to extra-dimensional models of flavor (see Ref.~\cite{vonGersdorff:2013rwa} for an overview of the constraints), with the important difference that here only fermionic resonances contribute to the effective operators, which strongly suppresses certain observables such as those from neutral meson mixing.

\subsection{Lepton observables}

\subsubsection{Decay $\mu\to e\gamma$}

This proceeds via the dipole operators $\mathcal (O_{eA})^{ij}=F_{\mu\nu}\bar e^i_L\sigma^{\mu\nu}e^j_R$ 
with coefficient
\be
(C_{eA})_{ij}=\frac{v}{\sqrt{2}}(c_wC_{eB}+s_wC_{eW})_{ij}\sim \eps_{\ell^i}\eps_{e^j}\frac{ec_e^3}{16\pi^2}\frac{v}{\mc_{e,\ell}^2}\,.
\ee
In terms of this, the partial width is
$
\Gamma(\mu\to e\gamma)=\frac{m_\mu^3}{4\pi}\left(\left| (C_{eA})_{12}\right|^2+\left|(C_{eA})_{21}\right|^2\right)
$. Using $\Gamma(\mu\to e\nu\bar\nu)=\frac{m_\mu^5}{384\pi^3v^4}$
one gets for the branching ratio 
\be
B(\mu\to e\gamma)
\sim 1.7\cdot 10^{-6} c_e^4 \left(\frac{{\rm TeV}}{\mc_{e,\ell}}\right)^{4}\left(\frac{\eps_{\ell^1}/\eps_{\ell^2}}{0.25}\right)^2\,,
\ee
where we have eliminated the $\eps_{e^i}$ parameters in favor of the physical Yukawa couplings $y_{e^i}\sim c_e \eps_{\ell^i}\eps_{e^i}$.
From the experimental bound $B(\mu\to e \gamma)<4\cdot 10^{-13}$ \cite{TheMEG:2016wtm} one finds that the $c_e\sim 1$ scenarios requires clockwork scales of the order of $\sim 40$ TeV, while the 2HDM scenario with $\tan\beta\approx 40$ is compatible with leptonic clockwork scales of only $1$ TeV. The  bounds are roughly consistent with what one finds in the analogous extra-dimensional models \cite{vonGersdorff:2013rwa}.

\subsubsection{$\mu\to e$ conversion}

The conversion of muons to electrons in nuclei such as gold is another strongly constrained observable.  In terms of the Wilson coefficients\footnote{We are neglecting QCD running of the 4-fermion operators between the weak scale and the $\mu$ mass for simplicity.} one has for the transition rate normalized to the capture rate \cite{Cirigliano:2005ck}
\be
B_{Au}(\mu\to e)=\frac{4 m_\mu^5}{\Gamma_{Au}^{\rm capt}}(|C_{H\ell}|^2+|C_{H e}|^2)\left|(1-4s_w^2)V_{Au}^{(p)}-V_{Au}^{(n)}\right|^2\,,
\ee
where the last factor comes from the coupling of the $Z$ boson to the nucleus.
Using the overlap integrals and capture rate from Ref.~\cite{Kitano:2002mt}
\be
V_{Au}^{(p)}=0.1\,,\qquad
V^{(n)}_{Au}=0.1\,,\qquad
\Gamma_{Au}^{\rm capt}=9\cdot 10^{-18} {\ \rm GeV}\,,
\ee
one obtains
\be
B_{Au}(\mu\to e)=13.1
\times c_e^4\left(\frac{v^4}{\mc_\ell^4}(\eps_{e^1}\eps_{e^2})^2
+\frac{v^4}{\mc_e^4}(\eps_{\ell^1}\eps_{\ell^2})^2\right)\,.
\ee
The experimental bound is $B_{Au}(\mu\to e)<7\cdot 10^{-13}$ \cite{Bertl:2006up}. 

Let us consider some representatives of the scenarios A, B as well as the 2HDM separately. In the following, we eliminate the $\eps_{e^i}$ in favor of the physical masses via the relation $y_{e^i}\sim c_e\eps_{e^i}\eps_{\ell^i}$, and use as reference values for the $\eps_{\ell^i}$ the  median values of the distribution in each case.

\begin{itemize}
\item Scenario A
($c_e=1$, $\xi_\ell=13$, $N_\ell=2$). 
\bea
B&\sim& 
5.8\cdot 10^{-9} \left(\frac{{\rm TeV}}{\mc_{\ell}}\right)^{4}\left(\frac{0.0008}{\eps_{\ell^1}}\right)^2
\left(\frac{0.006}{\eps_{\ell^2}}\right)^2\nn\\
&+&
9.8\cdot 10^{-13}\left(\frac{{\rm TeV}}{\mc_{e}}\right)^{4}\left(\frac{\eps_{\ell^1}}{0.0008}\right)^2
 	\left(\frac{\eps_{\ell^2}}{0.006}\right)^2.
\eea
\item Scenario B 
($c_e=1$, $\xi_\ell=1$, $N_\ell=2$).
\bea
B&\sim& 
5.2\cdot 10^{-17} \left(\frac{{\rm TeV}}{\mc_{\ell}}\right)^{4}\left(\frac{0.1}{\eps_{\ell^1}}\right)^2
\left(\frac{0.5}{\eps_{\ell^2}}\right)^2\nn\\
&+&
9.7\cdot 10^{-5}\left(\frac{{\rm TeV}}{\mc_{e}}\right)^{4}\left(\frac{\eps_{\ell^1}}{0.1}\right)^2
 	\left(\frac{\eps_{\ell^2}}{0.5}\right)^2.
\eea
\item 
2HDM scenario
($c_e\sim 0.025$, $\xi_\ell=1$, $N_\ell=4$).
\bea
{B}&\sim &
1.2\cdot 10^{-15} \left(\frac{{\rm TeV}}{\mc_{\ell}}\right)^{4}\left(\frac{0.03}{\eps_{\ell^1}}\right)^2
\left(\frac{0.35}{\eps_{\ell^2}}\right)^2\nn\\
&+&
2.0\cdot 10^{-12}\left(\frac{{\rm TeV}}{\mc_{e}}\right)^{4}\left(\frac{\eps_{\ell^1}}{0.03}\right)^2
 	\left(\frac{\eps_{\ell^2}}{0.35}\right)^2.
\eea
\end{itemize}

Keeping in mind the experimental limit, scenario B is the most constrained case, pointing to $\mc_e$  in the $\gtrsim 100$ TeV region, while scenario A requires $\mc_\ell\gtrsim 10$ TeV. The 2HDM scenario is the least constrained one, and survives even for TeV scale gear masses.

\subsubsection{Decay $\mu\to 3e$}

The branching ratio for the decay $\mu\to 3 e$ is governed by the exact same combination of Wilson coefficients as the $\mu\to e$ conversion rate. However, the bounds turn out to be about a factor of 3 weaker than the latter, hence we will not detail them here.

\subsection{Quark observables}

\subsubsection{Neutral meson mixing.}

Neutral meson mixing data impose some of the strongest constraints on New Physics. For the $K\bar K$ mixing, the relevant operators are $\mathcal O_{qq}^{(1,3)}$, $\mathcal O_{dd}$, and $\mathcal O_{qd}$. 
In our model, these coefficients are estimated in Tab.~\ref{tab:4f}, when evaluating them below, we simply take the median values of the $\eps_{\psi^i}$ parameters in every case, and also for simplicity choose $\mc\equiv \mc_u\sim\mc_d\sim \mc_q$. 
 
To find the limits, we perform a simplified analysis using the bounds on Wilson coefficients given in Ref.~\cite{Bona:2007vi,Isidori:2010kg}. 
In terms of the standard terminology,  $C^{sd}_1\sim (C_{qq}^{(1)}+C_{qq}^{(3)})^{2121}$, $\widetilde C^{sd}_1\sim (C_{dd})^{2121}$, and $C_{5}^{sd}\sim (C_{qd})^{2121}$.\footnote{The remaining coefficients, $C_{2,3}$, $\widetilde C_{2,3}$, and $C_4$ are not generated.}
The imaginary parts of these  Wilson coefficients are constrained as $C^{sd}_1,\widetilde C^{sd}_1\lesssim (1.7\cdot 10^{4}\ \rm TeV )^{-2}$ and $ C^{sd}_5\lesssim (1.4\cdot 10^{5}\ \rm TeV )^{-2}$ \cite{Bona:2007vi,Isidori:2010kg}. 

For clockwork scales $\mc\lesssim 4 \pi v\sim 3$ TeV, the  $Z$ exchange  between two $\Delta F=1$ vertices can actually generate the dominant $\Delta F=2$ effect, as the additional $(v/\mc)^2$ is smaller than a loop factor. 
This can be accounted for by multiplying the below estimates by $(4\pi v/\mc)^2$. 

\begin{itemize}
\item
Scenario A gives
\bea
C_1^{sd}&\sim& \left(  [8.5,6.2,6.3]\cdot 10^{4}\,   
\mc \right)^{-2},
\nn\\
\widetilde C_1^{sd}&\sim& \left(  [0.2,1.8,5.4]\cdot 10^{6}\,  
 \mc \right)^{-2},
\nn\\
C_{5}^{sd}&\sim& \left(  [1.2,3.4,5.9]\cdot 10^{5}\, 
 \mc \right)^{-2},
\eea
where the three values correspond to the three cases of scenario A in Tab.~\ref{TableQuark1}.
\item
The three cases of scenario B give
\bea
C_1^{sd}&\sim& \left(  [4.2,1.5,1.3]\cdot 10^{5}\,   
\mc \right)^{-2},
\nn\\
\widetilde C_1^{sd}&\sim& \left(  [9.1,6.0,31]\cdot 10^{5}\,  
 \mc \right)^{-2},
\nn\\
C_{5}^{sd}&\sim& \left( [6.2,3.1,6.4]\cdot 10^{5}\, 
 \mc \right)^{-2}.
\eea

\item
In the 2HDM scenario  one has
\bea
C_1^{sd}&\sim& \left(  3.7\cdot 10^{5}\,  
\mc
\right)^{-2},
\nn\\
\widetilde C_1^{sd}&\sim& \left(  2.1\cdot 10^{6}\, 
\mc
\right)^{-2},
\nn\\
C_{5}^{sd}&\sim& \left(  9.0\cdot 10^{5}\, 
\mc \right)^{-2}.
\eea
\end{itemize}
Clearly,  none of these cases result in any significant bounds on $\tilde m$ (say, above 1 TeV).

The bounds from the $D\bar D$  and $B\bar B$ systems are even weaker and will not be investigated here.

\subsubsection{Rare meson decays.}

Various measurements on rare decay modes of mesons, in particular $s$ and $b$ flavoured ones, constrain the tree-level generated flavor-violating $Z$ couplings.
We investigate here two of the experimentally best measured ones  that allow us to obtain the strongest bounds \cite{Tanabashi:2018oca}
\be
B(K_L\to \mu^+\mu^-)=(6.84\pm 0.11)\cdot 10^{-9},
\ee
\be
B(B_s\to\mu^+\mu^-)=(2.7\pm 0.6)\cdot 10^{-9},
\ee
which correspond to $s\to d$ and $b\to s$ transitions respectively. 

It is straightforward to relate the expressions in Ref.~\cite{Agashe:2013kxa} to the relevant Wilson Coefficients.
\be
B(K_L\to \mu^+\mu^-)=6.7\cdot 10^{-9} (1+[5911{ \rm\ TeV}^2 \operatorname{Re}(C_{Hq}^{(1)}+C_{Hq}^{(3)}-C_{Hd})^{12}-0.55]^2),
\ee
\be
B(B_s\to\mu^+\mu^-)=3.3\cdot 10^{-9}\left|1+70.1 { \rm\ TeV}^2 (C_{Hq}^{(1)}+C_{Hq}^{(3)}-C_{Hd})^{23}\right|^2.
\ee

In the following, we neglect the NP$^2$ terms, which, consistent with the EFT approach, are subleading.
Our class of models then yield
\begin{itemize}
\item Scenario A
\bea
B(K_L\to \mu^+\mu^-)&=&\left(7.0\pm [0.5,1.2,1.2]\left[\frac{\rm TeV}{\mc}\right]^2\right)\cdot 10^{-9},\\
B(B_s\to\mu^+\mu^-)&=&\left(3.3\pm [46,17,17]\left[\frac{\rm TeV}{\mc}\right]^2\right)\cdot 10^{-9},
\eea
where the $\pm$ schematically parametrizes the possible types of interference with the SM. For the second and third models the bounds are more severe than the $\Delta F=2$ constraints, going as high as 5 TeV for the third model and 9 TeV for the first one.
\item Scenario B
\bea
B(K_L\to \mu^+\mu^-)&=&\left(7.0\pm [0.10,0.38,0.58]\left[\frac{\rm TeV}{\mc}\right]^2\right)\cdot 10^{-9},\\
B(B_s\to\mu^+\mu^-)&=&\left(3.3\pm [8.2,4.0,10]\left[\frac{\rm TeV}{\mc}\right]^2\right)\cdot 10^{-9}.
\eea
The bounds are noticeably weaker than in scenario A, going maximally to 4 TeV in the case of the third model.

\item 2HDM scenario
\bea
B(K_L\to \mu^+\mu^-)&=&\left(7.0\pm 0.04\left[\frac{\rm TeV}{\mc}\right]^2\right)\cdot 10^{-9}\,,\nn\\
B(B_s\to\mu^+\mu^-)&=&\left(3.3\pm 0.05\left[\frac{\rm TeV}{\mc}\right]^2\right)\cdot 10^{-9}\,.
\eea
No significant bounds arise in this model, which can be consistent with data even for sub TeV masses.
\end{itemize}

\section{Conclusions}
\label{sec:conclusions}

In this paper we have proposed a model of flavor, both for the quark and lepton sectors, in which hierarchical masses and mixings arise from a chain of vectorlike fermions. In contrast to similar constructions with site-independent couplings, here the couplings at different sites are uncorrelated. We showed that if they  are taken to be random order-one complex numbers (in units of some  scale), it is very likely that the physical Yukawa eigenvalues of each species turn out hierarchical, thus explaining in a natural way the observed mass patterns in Nature.
We have also shown that this effect can be interpreted as a spontaneous and narrow localization of the zero mode at a random site of the chain ("localization in theory space"), which can occur close to the boundary where the Higgs lives (we identify this mode with a third generation fermion), towards the opposite site (first generation), or randomly in the bulk (second generation).
The deterministic parameters of the model are essentially the number of clockwork fermions of each gauge representation (in the case of $SU(5)$ compatible choices these are just two integer numbers), whereas effectively one more continuous parameter is needed in order to suppress the bottom and tau mass compared to the top mass. This latter parameter is of the order $\sim 10$, and can appear either as a ratio of scales or a $\tan\beta$ type parameter  in a 2HDM version of the model.
We have performed a quantitative analysis of the model by simulating the "posterior" distributions for the masses and mixings from reasonable prior distributions for the Lagrangian parameters, and show that for appropriate number of clockwork fermions  the experimental values appear near the mean values of these distributions, about one standard deviation away from it. The results are summarized in Tabs.~\ref{TableQuark1} and \ref{tab:lep}.

Furthermore, we have computed the dimension-six flavor changing operators (exactly at the tree level, and parametrically up to order-one numbers at the loop level). 
With these result in hand, we have investigated the main flavor changing effects and found that the primary constraints come from $\mu\to e $ conversion in nuclei, followed by $\mu\to e\gamma$. The constraints in the quark sector are less severe, and predominantly come from rare meson decays. The actual lower bounds on the vector-like scale differ quite a lot over the different variants of the model and can generally vary in the 1  to 100 TeV range.

We close with a few comments on possible further directions. It would be interesting to embed this paradigm in low-energy supersymmetry. Even if the flavor scale is very high, interesting bounds could be obtained due to the inheritance of the flavor structure by the soft terms. It has been pointed out previously \cite{Dudas:2010yh} that these effects can be markedly different for different flavor mechanisms (such as extra dimensions vs.~horizontal symmetries).
Furthermore, even though we have obtained good fits with $SU(5)$ compatible choices for the number of vectorlike fermions, the other Lagrangian parameters (the matrices $M_i^\psi$ and $K^{\psi}_i$) need to present a significant amount of  $SU(5)$ violation in  order to generate the differences in down and lepton Yukawa couplings.
However, interestingly, these couplings are dimension-3 operators, and hence they could descend from renormalizable Yukawa-type interaction with the GUT breaking {\bf 24} representation, which naturally can result in order one $SU(5)$ breaking effects. It would be very interesting to find a working model along these lines.

\section*{Acknowledgements}
GG wishes to thank the Conselho Nacional de Desenvolvimento Cient\'ifico e Tecnol\'ogico (CNPq) for support under fellowship number 307536/2016-5. We would like to thank Enrique Anda for useful discussions.

\appendix

\section{Masses and mixings: experimental data.}

\begin{table}[!h]
\centering
\subfloat[htc][Experimental data for the quark sector.]{\begin{tabular}{cc}
	\toprule
	 Quantity & Experimental value\\
	\midrule
	 $y_{d}$ & $1.364\cdot 10^{-5}$ \\
	 $y_{s}$ & $2.70\cdot 10^{-4}$\\
	 $y_{b}$ & $1.388\cdot 10^{-2}$\\
	 \midrule
	 $y_{u}$ & $6.3\cdot 10^{-6}$\\
	 $y_{c}$ & $3.104\cdot 10^{-3}$\\
	 $y_{t}$ & $0.8685$\\
	\midrule
	 $\sin{\theta_{12}}$ & $0.225$\\
	 $\sin{\theta_{23}}$ & $0.042$\\
	 $\sin{\theta_{13}}$ & $3.55 \cdot 10^{-3}$\\
	 $J$ & $3.18\cdot 10^{-5}$\\
	\bottomrule
	\end{tabular}}
	    \qquad \qquad
\subfloat[htc][Experimental data for the lepton sector.]{\begin{tabular}{cc}
	\toprule
	 Quantity & Experimental value\\
	\midrule
	 $y_{e}$ & $2.8482\cdot 10^{-6}$\\
	 $y_{\mu}$ & $6.0127\cdot 10^{-4}$\\
	 $y_{\tau}$ & $1.02213\cdot 10^{-2}$\\
	 \midrule
	 $\Delta m^{2}_{12}$ & $7.53 \cdot 10^{-5}$  $\text{eV}^{2}$ \\
	 $\Delta m^{2}_{23}$ & $2.51 \cdot 10^{-3} $  $\text{eV}^{2}$ \\
	 \\
	\midrule
	 $\sin^{2}\left(\theta_{12}\right)$ & $0.307$\\
	 $\sin^{2}\left(\theta_{23}\right)$ & $0.417$ \\
	 $\sin^{2}\left(\theta_{13}\right)$ & $2.12 \cdot 10^{-2}$\\
	 $J_{CP}$ & $-0.027$\\
	\bottomrule
	\end{tabular}}	
\caption{Experimental values for the masses and mixings \cite{Antusch:2013jca,Tanabashi:2018oca}. Yukawa couplings are evaluated at 1 TeV.
Experimental uncertainties are omitted as they are negligible compared with the widths of the theoretical distributions.}
\label{tab:exp}
\end{table}

\section{Random localization of the zero mode.}
\label{sec:anderson}

We are going to consider the CW model of one generation, Eq.~(\ref{eq:cw1}) with zero mode wave function $f_i$ defined as in Eqns.~(\ref{eq:wf0}) and (\ref{eq:eps}).

We are interested in the statistical distribution of the $f_i$ when $k_i$ and $m_i$ are randomly chosen from some given prior distributions. We will chose these priors (though not the actual values of $m_i$ and $k_i$) to be site-independent. 
If, in addition, the priors for $m$ and $k$ are identical, then the distribution for $x\equiv \log|k/m|^2$ is symmetric.
\be
p_x(x)=p_x(-x), \qquad x\equiv 2\log|k/m|\,.
\ee
For instance, for
uniform priors with ranges
\be
-1\leq m_i\leq 1\,, \qquad -1\leq k_i\leq 1\,,
\ee 
one has
\be
p(x)  = \frac{1}{2}\exp(-|x|)\,,
\label{eq:meas}
\ee
where $x$ runs over the entire real line.

For the moment we will keep the distribution $p(x)$ arbitrary.
Let us suppose that the wave function is maximized at site $j$,
$f_{\rm max}\equiv \max_i |f_i|= |f_j|$.
The set of values of the $x_i$ that corresponds to this situation defines a region in $\mathbb R^N$ that we will denote by $R_{N,j}$.
Using
\be
2\log \frac{|f_i|}{|f_0|}=\sum_{k=1}^i x_k\,,
\label{eq:log}
\ee
it is not hard to see that this region is defined by the relations
\be
x_j>0\,,\qquad  x_j+x_{j-1}>0\,,\qquad\dots\qquad x_j+\dots+x_{1}>0\,,\qquad 
\label{eq:r1}
\ee
and
\be
x_{j+1}<0\,,\qquad  x_{j+1}+x_{j+2}<0\,,\qquad \dots\qquad x_{j+1}+\dots +x_{N}<0\,.
\label{eq:r2}
\ee
Integrating  over $R_{N,j}$  then gives $\vol (R_{N,j})\equiv p_{N,j}$, the probability for the zero mode wave function to peak at site $j$. 
However, the relations in (\ref{eq:r1}) only involve $x_1\dots x_j$ and the ones in  (\ref{eq:r2}) 
only the $x_{j+1}\dots x_N$. 
In fact separately they define the two lower-dimensional regions $R_{j,j}$ and\footnote{Strictly speaking this notation implies a renaming $x_{j+i}\equiv y_{i}$ for $i\geq1$.} $R_{N-j,0}$ into which $R_{N,j}$ factorizes, in particular
\be
\vol(R_{N,j})=\vol(R_{j,j})\vol(R_{N-j,0})\,.
\label{eq:volfac}
\ee
Eq.~(\ref{eq:volfac}) then leads to the recursive relation
\be
p_{N,j}=p_{j,j}\,p_{N-j,0}\,.
\ee
If, in addition $p(x)=p(-x)$, one has the relation 
\be
p_{N,j}=p_{N,N-j}\,,
\ee
in particular 
\be p_{N,j}=p_{j,0}p_{N-j,0}\,.
\label{eq:P}
\ee
Hence, all probabilities are determined once we know the quantities $p_{j,0}$.
The latter can simply be obtained from normalization $\sum_{j=0}^N p_{N,j}=1$ and the relations Eq.~(\ref{eq:P}). One finds
\be
\qquad p_{j,0}=\frac{(2j-1)!!}{(2j)!!}=2^{-2j+1}{\binom{2j-1}{j}}\,.
\label{eq:pj0}
\ee
This result is independent of $p(x)$, as long as $p(x)=p(-x)$.

Next, we would like to investigate the conditional probabilities (given the location of the peak) that the wave function decreases say to the right of the peak. Formally, it is calculated as
\bea
p_-(k)\equiv \prob\bigl(|f_{j+k}|< |f_{j+k-1}|\,\bigm | f_{\rm max}=|f_j|\bigr)&=&\frac{\vol(R_{N,j}\cap x_{j+k}<0)}{\vol(R)}\nn\\
 &=&\frac{\vol(R_{N-j,0}\cap x_{j+k}<0)}{\vol(R_{N-j,0})}
\,.
\eea
In the second line we used the factorization fo the two regions which still holds true in the presence of the additional constraint $x_{j+k}<0$, which causes a cancellation of $\vol(R_{j,j})$ factors between numerator and denominator.
Notice that  the problem reduces from an $N$-dimensional to an $N-j\equiv N'$ dimensional one. Renaming the sites $x_i\to x_{i-j}$ we can simply calculate 
\be
p_-(k)=\prob\bigl(|f_{k}|< |f_{k-1}|\,\bigm| f_{\rm max}=|f_0|\bigr))= \frac{\vol(R_{N',0}\cap x_{k}<0)}{\vol(R_{N',0})} \,,
\ee
for $1\leq k\leq N'$.
Clearly $p_-(1)=1$ (otherwise $f_0$ would not be maximal).
We will now show that this probability decreases monotonically until it reaches 
the final value $p_-(N')=N'/(2N'-1)>\frac{1}{2}$. In other words, $p_-$ is always greater than $\frac{1}{2}$ and hence it is always more likely that the wave function decreases rather than increases at any step to the right of its peak.
Monotonicity is fairly straightforward. Let $R'=R_{N',0}$ and $R'_{i}\supset R'$ be defined like $R'$ but with the condition $x_1+\dots+x_i<0$ removed. Then
\bea
\vol(R'\cap x_{k}<0)&=&\vol(R'\cap x_{k-1}<0\cap x_k<0)+\vol(R'\cap x_{k-1}>0\cap x_k<0)\nn\\
&\leq& \vol(R'\cap x_{k-1}<0\cap x_k<0)+\vol(R'_{k-1}\cap x_{k-1}>0\cap x_k<0)\nn\\
&=& \vol(R'\cap x_{k-1}<0\cap x_k<0)+\vol(R'_{k-1}\cap x_{k-1}<0\cap x_k>0)\nn\\
&=& \vol(R'\cap x_{k-1}<0\cap x_k<0)+\vol(R'\cap x_{k-1}<0\cap x_k>0)\nn\\
&=&\vol(R'\cap x_{k-1}<0)\,.
\eea
In the third line we have used that the conditions in $R'_{k-1}$ as well as the integration measure are symmetric under exchange of $x_{k-1}\leftrightarrow x_k$, and in the fourth  line we used  
the condition $x_k>0\land x_1+\dots+x_k<0\Rightarrow x_1+\dots+x_{k-1}<0$.

In order to compute $p_-(N')$ we need
\be
\vol(R_{N',0}\cap x_{N'}<0)=\vol( x_{N'}<0)\vol(R_{N'-1,0})\,,
\ee
which follows from $x_{N'}<0\land x_1+\dots+x_{N'-1}<0\Rightarrow x_1+\dots+x_{N'}<0$, such that the latter condition can be dropped without consequences. By the use of Eq.~(\ref{eq:pj0}) one finds
\be
p_-(N')=\frac{\vol(R_{N',0}\cap x_{N'}<0)}{\vol(R_{N',0})}=\frac{\frac{1}{2}p_{N'-1}(0)}{p_{N'}(0)}=\frac{N'}{2N'-1}\,,
\ee
thus establishing the result.

\section{Mass eigenbasis}
\label{sec:mass}

It is useful to compute the Wilson Coefficients in the mass eigenbasis before electroweak breaking, which is less explicit than the basis employed in the main text but nevertheless insightful.
The two type of operators are then generated by tree-level diagram as shown in Fig.~\ref{fig:tree}.
To avoid an overboarding notation, it is useful to first consider the case of a single generation.
For instance, the  diagram involving exchange of up type gears generates the unique dimension-six operator
\be
\mathcal L_{6,u}=
\bar q_L\tilde H \, (U^{q_L})^*_{0,0} \, C_u \left(\sum_{a\neq 0}\frac{|U^{u_R}_{0,a}|^2}{(m^{u}_{a})^2}\right)
 C^*_u\, U^{q_L}_{0,0} (i\Dsl)( \tilde H^\dagger q_L)\,.
\label{eq:dim6mass}
\ee
After reduction to the Warsaw basis, we end up again with the previous two type of operators of Tables \ref{tab:yuk} and \ref{tab:cc}, but this time expressed  in terms of the mass eigenvalues and mixing angles.
The notation is as follows: $U^{q_L}$  ($U^{q_R}$) are defined as the $N_{q}+1$  ($N_{q}$) dimensional unitary matrices diagonalizing the mass matrix for the doublet gears, and $U^{u_R}$ and $U^{u_L}$ (of dimension $N_{u}+1$ and $N_{u}$ respectively) are the analogous matrices for the singlet gears.
Notice that the first column of $U^{q_L}$ contains the zero mode wave function,
\be
U^{q_L}_{i,0}=f^{q_L}_{i}\,,
\ee
in particular 
\be
U^{q_L}_{0,0}=f^{q_L}_{0}=\eps_q\,.
\ee
Furthermore, $a$ labels the $N_u$ mass eigenstates.
 The term in parenthesis in Eq.~(\ref{eq:dim6mass}) is essentially a weighted average over the inverse gear masses squared. To simplify this expression, we make the approximation of approximately degenerate gear masses, that, is 
\be
 m^u_a\sim \mc_u\,,
 \label{eq:approxcw}
\ee 
where $\mc_u$ is some characteristic clockwork scale for the $u$ sector. 
A possible definition for this scale is given in Eq.~(\ref{eq:defmc}).

Unitarity implies $\sum_{a\neq 0}|U^{u_R}_{0,a}|^2= 1-\eps_u^2$ and  the approximation Eq.~(\ref{eq:approxcw}) then simplifies the dimension-6 Lagrangian in Eq.~(\ref{eq:dim6mass}) as
\be
\mathcal L_{6,u}=
\frac{\eps_q^2(1-\eps_u^2)|C_u|^2}{\mc_u^2}\
\bar q_L\tilde H
(i\Dsl)( \tilde H^\dagger q_L)\,.
\ee
Similar operators appear from the exchange of the other 4 types of gears.

When going to three generations, $U^{q_L}_{0,0}$ becomes a 3 by 3 matrix, previously denoted by $\E_q$, see Eq.~(\ref{eq:Eps}).
Moreover, the single row matrix $U^{u_R}_{0,a}$ becomes a $3\times 3N_u$ matrix that we will denote by $G_{u_R}$. Then for instance, one has
\be
C_{Hq}^{(1)}=\frac{1}{2}\E_q C_u G_{u_R}  \mathcal M^{-2}_{u} (G_{u_R})^\dagger C_u^\dagger \E_{q}+{\rm exchange\ of\ down\ gears} \,,
\ee
where in our convention the $\E$ matrices are Hermitian, and we defined the diagonal $3N_u\times 3 N_u$ matrix $\mathcal M_{u}$ that contains the eigenvalues $m^u_a$. Notice that the three rows of $(\E_u,G_{u_R})$ 
comprise the first three rows of the unitary matrix $U^{u_R}$.
Comparison with Tab.~\ref{tab:cc} gives
\be
\E_{u}A_{u}\E_{u}=G_{u_R}  \mathcal M^{-2}_{u} (G_{u_R})^\dagger\,.
\label{eq:rewrite}
\ee
Note that all tree-level Wilson Coefficients in Tab~\ref{tab:yuk} and \ref{tab:cc} contain the combination $\E_{\psi}A_{\psi}\E_{\psi}$, and in analogy to Eq.~(\ref{eq:rewrite}) one has the relations
\be
\E_{\psi}A_{\psi}\E_{\psi}=G_{\psi}  \mathcal M^{-2}_{\psi} (G_{\psi})^\dagger\,.
\label{eq:rewrite2}
\ee
Even though we do not have analytical expressions for the masses and mixings on the right hand side of Eq.~(\ref{eq:rewrite2}), this is still a useful rewriting. For instance, approximating the eigenvalues by the CW scale $\mc_u$ and using unitary $\E_\psi^2+G_{\psi}(G_{\psi})^\dagger=\one_{3\times 3}$ one gets
\be
\E_{\psi}A_{\psi}\E_{\psi}\approx \frac{1}{\mc_\psi^2}G_{\psi}(G_{\psi})^\dagger=\frac{1}{\mc_\psi^2}( \one -\E_\psi^2).
\ee
One possible definition for the CW scale $\mc_\psi$ is in terms of the prior distributions for the matrices $M_\psi$ and $K_\psi$. If we denote the medians of the $|(M_\psi)_{ij}|$ by $\bar m_\psi$ and the medians of $|(K_\psi)_{ij}|$ by $\bar k_\psi$, we define
\be
\mc_\psi^2=\bar m_\psi^2+\bar k_\psi^2\,.
\label{eq:defmc}
\ee
In practice, this defines the scale that sets the mass eigenvalues of the clockwork gears.

\bibliographystyle{JHEP}

\bibliography{paper}

 
\end{document}